%% file: GDP.tex
\documentclass[10pt]{article}
\usepackage[a4paper, bottom=1cm]{geometry}
\usepackage{endnotes}
\usepackage{graphicx}
\usepackage{fullpage}
\usepackage{setspace}
\usepackage{lineno}
\usepackage{url}
\usepackage{subfigure}
\usepackage{epstopdf}
\usepackage{cite}
\usepackage{graphicx}
\usepackage{mdwtab}
\usepackage{mdwmath}
\usepackage{multicol}
\usepackage{amssymb, amsmath}
\usepackage{color}
\usepackage{txfonts}
\usepackage{CJK}
\usepackage[official]{eurosym}
\begin{document}
\title{Does GDP measure growth in the economy or simply growth in the money supply?}
\author{Jacky Mallett and Charles Keen}
\date{1st August 2012}
\maketitle
\section*{Abstract}
Gross Domestic Product(GDP) is a widely used measurement of 
economic growth representing the market value of all final goods and services
produced by a country within a given time.  In this paper we question
the assumption that GDP measures production, and suggest that in reality
it merely captures changes in the rate of expansion of the money supply
used to measure the price data it is derived from. We first review 
the Quantity Theory of Money $MV=PT$, 
and show that the Velocity of Circulation of Money(V) does not affect the price level
as claimed,
as it is also a factor of the quantity of transactions(T). It then follows directly that
attempts to measure total production from any form of price data as the GDP measurement does, 
will necessarily be confounded
by the inverse relationship between prices and the quantity of production, 
which requires that as the total quantity of production increases, prices will drop.
Finally, in support of this claim
we present an empirical analysis of the GDP of nine countries and one currency union,
showing that when normalized for money supply growth
GDP measures have been uniformly shrinking over the last 20 years, and discuss the possible
reasons for this behaviour. 
%
\doublespacing
\section*{Introduction}
Gross Domestic Product(GDP) is an internationally used measurement of an economy's 
output and performance. It is formally defined as the market value of all final goods and services 
produced within a country in a given period. Originally developed by Dr. Simon Kuznets\cite{kuznets.1934}
for the United States Government 
in 1934, it is used as the basis for several derivative statistics, the 
Gross Domestic Product per 
Capita, a measure of a country's output per person, and real GDP which is the base GDP measurement 
adjusted for consumer price inflation(CPI). As described in the U.S. Department of Commerce's
primer on National Income and Product Accounts\cite{bea.2007}, GDP is one of a series
of measurements that provide "information on the value and composition of output produced
in the United States". GDP statistics are usually provided as a raw measurement referred to as the nominal
GDP, and real or constant GDP which is the nominal GDP adjusted for price inflation.
\par
In this paper we challenge whether even the inflation adjusted 'real' GDP statistic measures growth in the 
production of goods and services in the way commonly claimed.  
Although GDP is cited as being a measure of "economic growth", and is often thought of as
being a direct measurement of an economy's production of goods and services, it is
calculated from a variety of statistical and survey data that is derived from the
prices of the goods and services included in its scope. This implies that price data
can be used to extract quantitative information on the production of goods and services.
We argue that this is not possible because of the inverse relationship between the price level
and changes in the quantity of transactions of goods and services.
\par
Typically, GDP is used by economists for comparative measurements.
Press reports such as "India's economy grew at the slowest rate for more than
two years"\footnote{Source, Financial Times November 30th 2011, Headline "Indian GDP growth 
drops below 7\%"} are based on GDP figures, as are official reports on policy measures
and their outcomes.  For example,
in its third quarter bulletin in 2011 the Bank of England summarizes the macro-economic
effects of its recent quantitative easing program by saying that "QE may have raised the 
level of real GDP by 1$\frac{1}{2}\%$ to 2\% and increased inflation by $\frac{3}{4}$ to $1\frac{1}{2}$
percentage points"\cite{boe.2011}. 
Cross country comparisons within economic literature also frequently use GDP as the measure
for comparison. To take two of many possible examples,
Acemoglu\cite{acemoglu.2007} uses 
a multi-country examination of predicted mortality and GDP to report
that increases in life expectancy did not lead to increases in income per capita, and 
Afonso\cite{afonso.2012} uses a Bayesian Structural Vector Autoregression approach to
show that government spending shocks have a small effect on GDP, but appear to generate
a quick fall in stock prices.
\par
\par        
In a markert based economy price data is not an absolute number, but rather a continuously
changing relative measurement, that is determined both by the supply and demand for goods and services, and the
total amount of money available to mediate that exchange. Although we use 
money for measurement purposes when examining the economy its main role in the economy is as a token
of exchange, and units of money are repeatedly re-used over time for this purpose as they
circulate within the monetary system.  This relationship is expressed in Economic theory as the Quantity Theory of 
Money(\ref{eq:mvpt}):
\begin{equation}
MV = PT
\label{eq:mvpt}
\end{equation}
where M is the total quantity of money, V is defined as the velocity of circulation of money,
the number of times each unit of money is reused in the time period measured, P is the average
price level of all transactions, and T is the total number of all transactions performed. 
\par
This equation implies
that if the money supply is held constant, then the relationship between price and the number of transactions
varies with the velocity of circulation. However, as we will show
below, since T is the number of all transactions performed, it must also be
a multiple of V, and V then cancels. We are left with 
prices being inversely related to the number of transactions, intermediated by the total money supply. 
The number of times each unit of money circulates in the economy, somewhat counterintuitively, plays
no part in determining the price level.
The implication for GDP
is that an overall increase in production will cause a commensurate drop in prices, and so 
the total quantity of production cannot be obtained solely from the sum of its price data.
For example, if we take a simplistic model of pricing, and assume that as production quantity doubles, 
the price of each good halves, and that the entire money supply is being continuously used for transactions,
it follows that the sum of the prices of total production sold remains unchanged. 
Assuming that the prices of all goods exchanged were
included in the GDP measure it would consequently be constant regardless of changes in the underlying quantity of 
goods and services, as prices adjusted to different levels of supply and demand.
However money supply data for modern economies shows a consistent pattern of 
increasing money supplies, punctuated by occasional contractions, 
and has done for several centuries.
If this growth in money, is not correctly compensated for then GDP levels 
may appear to grow (or shrink), purely
as a result of the  change in the underlying unit of measurement, money.
\par
Statistically, three different methods exist to calculate GDP, which should all provide the same result,
the production approach, the income approach, and the expenditure approach.
The production approach, which is also known as the net product or value added method
is based on estimates of the gross value of domestic output, and sums the outputs of 
every class of enterprise that qualifies
for GDP inclusion, the income approach is derived primarily from data on salaries, farmers' income, 
and corporate profits, and the expenditure approach is based on estimates of the total amount
of money used to buy production. The series of examples described by the U.S Department of 
Commerce\cite{bea.2007} shows that calculation in all cases is based on statistical data derived
from estimates based on price or income data, for example the total sales of durable and non-durable
goods.  Landefeld\cite{landefeld.2008} 
also provides several examples showing the price data that GDP is calculated from.
The expenditure method for example uses the following formula to calculate GDP:
\begin{displaymath}
GDP = private consumption + gross investment + government spending + (exports - imports)
\end{displaymath}
all of which are obtained from measurements made using the currency of the economy
in question, and the price of goods and services in that currency.
\par
A further problem arises since not all monetary transactions are included in the GDP measurement.
Significantly the majority of monetary transactions conducted in
the financial sector, the sale of shares and bonds for example, are not counted in GDP, nor
are intermediate exchanges within the production chain - GDP is based on the value 
of final goods. Consequently the observed
behaviour of a GDP measurement normalized for any underlying money supply growth,
could also be expected to vary over time, if shifts occurred in the amount of spending on goods and
services that are included in GDP, and those that are not. With the noted growth in the
financial sector in many economies over the last 20 years, the majority of
whose transactions are not included in GDP, the long term profile of the GDP statistic for economies 
with significant
financial sectors would be predicted to be a steady decrease once
the effect of money supply growth on the unit of measurement had been removed. 
\par
In the rest of this paper, we will review the quantity theory of money, 
with respect to the relationship between price, production and
the velocity of circulation of money, and their relationship to GDP. We will then present empirical 
data obtained from 
a comparative survey of 9 single currency economies selected to be representative of countries with
relatively stable long term fiscal behaviour, and one currency union, the Euro.
Defining money as the sum of all physical currency and bank deposits (see below), 
GDP figures were normalized against their country's central bank's monetary measures
over the period of the survey. Results for each country are presented alongside the original
GDP figures and money supply figures. In each case, the behaviour of the GDP statistic
supports the hypothesis, and shows a generally decreasing level against the largest monetary measure for
each currency.
\par
This analysis also highlighted 
other issues with contemporary monetary statistics. The precise definition of the money supply
appears to be unclear with a wide range of quantitative measures
provided by different Central Banks. Significant variations were also found in 
the definitions of these measurements, with definitions varying between countries for 
commonly used aggregate statistics such as M2, M3, etc. Consequently we have 
included the complete definition of the monetary measures 
provided by the Central Bank for each currency.
\section{Quantity theory of Money}
The formulation of the quantity theory of money(\ref{eq:mvpt}) is generally 
attributed to Fisher\cite{fisher.1911} although the idea precedes him. 
Fisher published his theory in 1911, at which time money was generally defined by economists as 
physical currency, notes and coins, and the classification of bank deposits as money was disputed. 
Fisher's definition of money did not include bank deposits, and he states:
\begin{quote}
But while a bank deposit transferable by check is included as circulating media, it is not money. 
A bank note, on the other hand, is both circulating medium and money. Between these two 
lies the final line of distinction between what is money and what is not. True, the line is 
delicately drawn, especially when we come to such checks as cashier's checks or certified checks, 
for the latter are almost identical with bank notes. Each is a demand liability on a bank, 
and each confers on the holder the right to draw money. Yet while a note is generally 
acceptable in exchange, a check is specially acceptable only, i.e. only by the consent of the payee. 
Real money rights are what a payee accepts without question, because he is induced to do so 
either by legal tender laws or by a well-established custom.\footnote{The Purchasing Power of Money, Chapter 2 Section 1.}
\end{quote}
This discussion of the definition of money is significant, and not only in the context of its use
as a unit of measurement. In the normal operation of the banking system, 
bank deposits are created through the extension of bank loans, and removed as loan principal 
is repaid, and considerably larger amounts of money are represented as on deposit in bank accounts
than is physically circulating in the economy. However that bank deposits should be counted as part of
the money supply, and indeed also that the process of bank lending could cause increases 
in the amount of money
represented as bank deposits, was only beginning to be explored
by the economists of Fisher's era. It would not be partially acknowledged until the 
1931 Macmillan report\cite{macmillan.1931} which included a description of the loan/re-deposit
process embedded in banking which appears to have been authored by Keynes\cite{stamp.1931}. 
Without a full understanding of this phenomena, and a definition
of money that excluded the majority of the active money supply\footnote{Estimates from 
the 1890's were that over 90\% of all financial transactions were then being made
directly between bank accounts using cheques and bills\cite{redman.1900}.} some form
of compensating component in Fisher's equation for a source of money supply expansion would be
required in order for early 20th century measurements of the money supply and price data to be
rationalized, and the intrinsically nebulous "Velocity of Money" appears to have fulfilled
this purpose.
\par
It can in fact be demonstrated using the examples that Fisher provides in the first chapter of his book, that 
the presence of V, 
the number of times individual units of money circulate in the economy in the equation is 
superfluous.  From Chapter 1:
\begin{quote}
"Let us begin with the money side. If the number of dollars in a country is 5,000,000, and their 
velocity of circulation is twenty times per year, then the total amount of money changing hands 
(for goods) per year is 5,000,000 times twenty, or \$100,000,000. This is the money side of the 
equation of exchange.  Since the money side of the equation is \$100,000,000, the goods side 
must be the same.  For if \$100,000,000 has been spent for goods in the course of the year, then 
\$100,000,000 worth of goods must have been sold in that year."
\begin{table}[ht]
\centering
\begin{tabular}{|l|c|ll|}
\hline
\$5,000,000 x 20 times a year& =  &    \multicolumn{2}{c|}{}      \\
 & +   &200,000,000 loaves of bread x &\$.10 a loaf \\
 & +   & 10,000,000 tons of coal    x &\$5 a ton    \\
 & +   &  30,000,000 yards of cloth x &\$1 a yard   \\
\hline
\end{tabular}
\caption{Example provided by Fisher in Chapter 1 for the Velocity of Circulation}
\label{velocity_eg1}     
\end{table}
\end{quote}
Examining this example more closely, the physical money supply is \$5,000,000, 
and consequently it is not possible to make a single purchase of the 10,000,000 tons of coal 
at \$5 a ton, which would require \$50 million. 
Transactions must proceed as a series of exchanges. If each transaction uses the entire available 
money supply, then assume
that the cloth makers buy 1 million tons of coal from the coal miners,
then the coal miners buy \$5 million of bread, and the bread makers then buy 5 million
yards of cloth, thereby allowing the cloth makers to buy the next 1 million tons of coal. In total this 
gives the following numbers of exchanges:
\begin{table}[ht]
\centering
\begin{tabular}{|l|c|c|l|ll|}
\hline
\$5,000,000 & V       &=&        & \multicolumn{2}{c|}{}                        \\
            & x 4     &+& 4  x   & 50,000,000 loaves of bread &x \$.10 a loaf   \\
            & x 10    &+& 10 x   &  1,000,000 tons   of coal  &x \$5   a ton    \\
            & x 6     &+& 6  x   &  5,000,000 yards  of cloth &x \$1   a yard   \\
\hline
\end{tabular}
\caption{Detailed breakdown of Exchanges shown in Table \ref{velocity_eg1}}
\label{velocity_eg2}     
\end{table}
\par
Each time a monetary purchase occurs there is necessarily
a matching transaction, and consequently V is present on both sides of the equation
and cancels\footnote{Other authors have raised this 
issue with velocity in various contexts, notably Lounsbury\cite{lounsbury.1931}
writing in 1931 conducted an extensive critique of both Fisher and Keynes' use of
it in their theories. Nevertheless several recent papers discuss its purported effects
including Mendiz\'abal's investigation of its lower than expected effect on inflation
suggesting a relationship between transaction technologies\cite{mendizabal.2006}, 
and Faig's investigation of the relationship between velocity and savings, which 
explicitly assumes that inflation
increases the velocity of circulation of money\cite{faig.2007}}.
\par
Using the same example to illustrate the problem with measuring GDP, let bread and cloth be classified as final goods 
for the purposes of calculating the 
GDP, and coal as an intermediary good which is not included in the measurement. The GDP in Fisher's simple 
economy is then \$50,000,000. If we 
double the quantity of all production in Table \ref{velocity_eg1}, and 
recalculate the price from m, V and t, GDP remains \$50,000,000 as the doubling in production has 
caused all prices to halve.
Increases in the velocity of circulation then cannot influence price, since they must occur
on both sides of the equation; but an increase in the quantity of money(m) will directly increase
the price, and through it GDP. GDP would also change if less coal were
sold, and more bread and cloth were purchased assuming their price increased as more money was 
switched to their purchase. This change would however be purely an artifact of some goods not being included
in the GDP measure, and so also not a measure of increased production. Indeed, if a fall in food production 
caused a shortage, and prices rose as a result of money being diverted from the purchase of goods that were not
included in GDP, then GDP would increase, even though the amount being produced of an economically
critical good had in fact dropped.
\begin{table}[ht]
\centering
\begin{tabular}{|l|c|c|l|ll|}
\hline
\$5,000,000 & V       &=&        &    \multicolumn{2}{c|}{}                      \\
            & x 4     &+& 4  x   & 100,000,000 loaves of bread &x \$.05 a loaf   \\
            & x 10    &+& 10 x   &   2,000,000 tons   of coal  &x \$2.5 a ton    \\
            & x 6     &+& 6  x   &  10,000,000 yards  of cloth &x \$0.5 a yard   \\
\hline
\end{tabular}
\caption{Increases production with associated recalculated prices.}
\label{velocity_eg3}     
\end{table}
\par
\section{Methodology}
The precise definition of money has long been a matter of debate within economic theory. 
Initially money was regarded as a purely physical token of metals of different rarities,
but as banking became more widespread, 
the status of bank deposits as money began to be debated. By the early 20th century 
as the preponderance of bank mediated transactions in the economy became clear, the discussion
within economics
was shifting from declarations such as Fisher's that bank deposits were not money, 
to that of which bank deposits should be classified as money. By the latter half of the century
a variety of formal measurements, M0, M1, M2, M3, M4, MZM, etc. had been introduced by
the national monetary authorities. 
\par
Within the different categories used for monetary measurement, there is a general convention
of low numbers indicating restricted definitions of money, with M0 or M1 typically
used to categorize physical money and instant access accounts, whilst
higher measures such as M3 and M4 introduce deposits with time access
requirements and various forms of debt. There is also a convention that higher monetary
aggregates include lower ones, although this is not invariably the case.
\par
From the viewpoint of economists one of the problems with simply classifying all bank 
accounts as money is the liquidity of the various types of bank accounts
offered by banks, with some accounts requiring periods of notice before funds can 
be accessed. The observation can be made that on aggregate, and especially for measurements
such as GDP which are computed over a timespan of several months, 
access to and the use of these forms of bank account
can be expected to average out over time. It can also be observed that their inclusion in the 
quantity of money does not raise any particular definitional issues, as when they are accessed the funds in
them function directly as tokens or units of exchange, and they are classified as liabilities on
bank's accounting sheets. More problematic is the presence of
money market funds in some monetary measures, such as the USA's M2 statistic. Although money market
funds are described as being equivalent to money, their funds are typically held in a mixture of
of deposit accounts and short term commercial and government debt. Their liquidation
by their depositors thus requires an exchange with money to occur, and it is  
incorrect to classify them simply as money.
\par
An additional issue with the various measures of the quantity of money is that they 
do not appear to have been standardized. There is considerable 
variation between countries not only in which measures are used, but also the components of the money
supply that are included in measures with the same name. For example the definitions of M2 used
by the USA for the dollar, and by the European Central Bank for the euro are
significantly different. In this paper we have used the sum of all physical money and bank deposits
as the guiding principle for measurement. Monetary measures have been selected that reflect this   
definition from the countries examined, with the lower measures also provided when applicable 
so that comparisons can be made
of the results of normalization against all the different measures available.
In practice for most countries the M2 or M3 definition of the money supply is the most
representative, although in the case of the USA adjustment was required in order to remove sub-components representing 
money market funds and other forms of debt. 
Long term time deposits are included in this definition of money, even though they are 
nominally not accessible without penalty. They represent money and are treated as liabilities 
in bank accounting. We believe that it is reasonable to assume for the purposes of this analysis,
that access and use of the money 
in these accounts is evenly distributed across the period of measurement.
\par
Normalization of the GDP measurement is performed with respect to the different measures
of the money supply at the beginning of the period 
being examined for the relevant M1, 2, 3 etc. component, using the formula:
\begin{displaymath}
Normalized\;GDP_{t} =  \frac{GDP_{t}}{1 + \frac{m_{t} - m_{0}}{m_{0} }}
\end{displaymath}
where $m_{t}$ is the relevant money supply component at time $t$. Both the normalized charts for
GDP, and the original data of the money supply and GDP are presented for each country.
\par
GDP data is taken in all cases from the International Monetary Fund's(IMF) September 2011 
World Economic Outlook Report. The series used are the Gross Domestic Product, current prices(National
Currency series) which is uncorrected for any inflationary effects. 
\par
Countries were chosen to be a representative sample both of different types of economy,
industrial, agricultural, financial, energy, etc. and of
different sizes, but with the requirement that their monetary history be sufficiently
stable to offer a reasonable basis for comparison, and that their statistical series be
publicly available. This excluded countries such as Brazil
for example, which has unfortunately had to reset its monetary system several times over the last decades.
The Euro is included since it represents to significant an economy to exclude,
but it has the shortest time series, and it can also be expected that its aggregate GDP
masks wide differences in national GDP within the currency union. 
\input{cds.tex}
\section{Conclusion}
Empirical evidence from the last 200 years, where available\footnote{Not all currencies
have a continuous history, and not all statistical series are available online. The
longest known online series, is that of the 
central bank of Sweden(Sveriges Riksbank) which provides a continuous series from 
1871(\url{http://www.riksbank.se/templates/Page.aspx?id=27395}).}
shows a more or less continuous expansion in the money supply for all currencies
during the reserve based banking era, when money is measured as the total of all physical notes
and coins together with bank deposits in the banking system. Absent occasional sharp contractions due to widespread
banking failure such as occurred in the Great Depression, the normal state of any
country's money supply is expansion, either due to regulatory failures within the banking
system itself, or deliberate injections into the system by the 
Government, and sometimes both.
\par
The old question of whether or not bank deposits should be classified equally 
with physical money, which is not entirely without merit when issues around 
time deposits are considered, has become increasingly irrelevant in systems where 
the majority of all transaction are carried out through electronic banking. The problem 
confronting economics is simply
that money is used as unit of measurement, but that its quantity is 
continuously changing at varying rates over time and place. While
the ability of the mechanical processes of the banking system to generate money are discussed
under the rubric
"Endogenous Money Expansion"\cite{lavoie.2003} within Economics, the 
full implications, particularly with respect to the interpretation of price based
statistics such as GDP do not appear to have been completely appreciated. As the normalized
charts for GDP in this paper demonstrate, had money supplies been constant over the periods
measured, the anomalous behaviour of production measurements based on price data would have been
apparent, instead it was effectively masked by the underlying growth in the
money supply. 
\par
Sharp criticisms of economics, and in particular macro-economic
theory have become commonplace in recent years, particularly as a result of the discipline's
general failure to either predict or explain the series of credit crises that
have so far dominated 21st century monetary systems. However, these criticisms generally
fail to provide concrete answers as to how an entire field of research could be
so poorly grounded, and it is unfair to researchers in the economic academy to 
make such an extraordinary claim, without also attempting to satisfy the associated burden of scientific proof. 
Unrecognized issues with a commonly used unit of measurement
such as money, would provide at least a partial explanation. They would also offer   
an answer to why work in macro-economics in particular has proved so problematic.
\par
As the branch of economics that deals with the structure,
performance and behaviour of the entire economy at regional, national and
global levels, macro-economics is also the area that is most exposed to these issues of monetary
measurement. Work in the field is generally based on the interpretation of a range of gross
statistical measurements, including GDP, inflation, unemployment, levels of
imports and exports, currency exchange rates, and private and governmental debt. 
With the exception of unemployment figures, all of these measurements are made from price data.
As demonstrated even with the relatively short periods used in this paper, rates of 
monetary expansion vary considerably both between countries, and also over time within
the same country, thus providing a confounding element for all forms of analysis based on 
monetary information. Attempts to remove this element by correcting data for CPI inflation ('real' measurements) 
unfortunately fails to incorporate the other factors
that acts on price data, changes in production and shifts between measured and unmeasured sectors. 
\par
The impact of these varying rates of monetary expansion over time and place would be in 
direct proportion to the length of the time period being examined, making macro-economic research particularly
vulnerable due to the longer time periods typically involved.  Additional 
issues would be introduced when comparisons were being made between countries or historical periods with
different rates of money supply growth.
\par
It should not be assumed that all results based on price data are invalid, approaches
that are based on calculating ratios between sectors would be less affected by some of these issues.
Some results may also be serendipitously correct. The use of GDP to debt ratios as gauges of fiscal
stability for example, since for many of the countries examined
here there appears to be a fairly close correspondence between the behaviour of total GDP and the larger
money supply measures. The ratio of the money supply to the total outstanding debt within an economy
is probably the most useful measure in gauging the likely default rate for any given country,
and the GDP ratios being used contain at least some of the required monetary information.
\par
The statistical problems with interpreting measurements based on money are certainly not confined to GDP. 
Table \ref{tab:govt_debt}\footnote{Source, IMF World Economic Outlook Database September 2011. Bank deposit information
for the Eurozone obtained from the OECD Bank Profitability Series, 1999-2009
\url{http://stats.oecd.org/index.aspx?DataSetCode=BPF1}} shows
the increase in Government debt for selected economies between 1999 and 2009, normalized
for the underlying growth in the money supply. Without normalization, the true picture
of government debt is hard to determine. Increases in the money supply over time will 
cause the gross amount of debt to increase, but whether this is proportionally a larger or smaller
amount of the economy can only be inferred through normalization. 
\begin{table}[ht]
\centering
\begin{tabular}{l|l|l|l|l|l}
Country   &  1999           &  2009                & Money Supply  & Normalized 2009     & \% increase \\
          &                 &                      & Multiplier    &                     &             \\
\hline
UK      &  \textsterling354.4  &  \textsterling759.5  & 2.6        & \textsterling292.12 & -17\%   \\
USA     &  \$5,662             & \$13,972             & 2.0        & \$6,986             & 23\%    \\
Germany &  \euro1,225          & \euro1,760           & 2.0(1.4)   & \euro880 (1257)     & -28\% (0\%)   \\
Spain   &  \euro361            & \euro561             & 2.0(3.0 )  & \euro281 (187)      & -22.0\% (-48\%)  \\
Italy   &  \euro1,281          & \euro1,763           & 2.0(2.0)   & \euro921            & -37\%   \\
\hline
\multicolumn{6}{l}{* Total Eurozone (Country's individual bank deposit expansion)}
\end{tabular}
\caption{Government debt levels normalized for money supply increases (Currency Unit Billions)}
\label{tab:govt_debt}
\end{table}
\par
Of particular note here is the behaviour of countries within the Eurozone, which show significantly
varying rates of internal expansion within their individual banking systems, that is not captured
by aggregate information on the Euro.  While government debt has
increased in all countries as an absolute amount, it has also shrunk in proportion to the total
money supply, leaving open to interpretation whether the economic problems in the eurozone are in 
fact a problem of gross government debt, excessive private sector debt competing with government debt,
problems with revenue caused by inconsistencies in taxation arising from a perception of debt growth, 
capital flows between countries, or indeed all of the above. 
All we can say with certainty from
the varying rates of monetary expansion within the Eurozone shown in the OECD figures, is that
the problems of the Euro are certainly not confined to government debt. 
\par
The issue of how economic growth or performance can usefully be measured remains open,
especially with respect to the confounding effects introduced by continuously adapting
prices, and shifts in the distribution of money within economic sectors. The problem
facing economic analysis is not only correctly identifying the quantity of money
and correcting for its changes, but also what reliable information if any can 
be extracted from price data? There is a distinct danger that until these issues are properly
recognized, policy initiatives will inadvertently favour interventions that lead
to money supply increases, since these will appear to increase "growth" as measured
by GDP. 
\par
Extensive analysis based on normalized measurements, and an accompanying review of all
results derived from GDP and other price related statistics is needed, before a clearer
picture can emerge of the overall effect of these distortions on macro-economic theory itself. 
It seems quite probable though, that with corrected statistics available, it will be possible
to form a considerably
clearer picture of the actual issues facing modern economies.
\include{biblio}

\end{document}

%% file: cds.tex
\section{Country Data Series}
\subsection{USA: 1980 - 2010}
\begin{figure}[ht]
\begin{minipage}[t]{7.5cm}
\begin{center}
\includegraphics[width=7.5cm, clip]{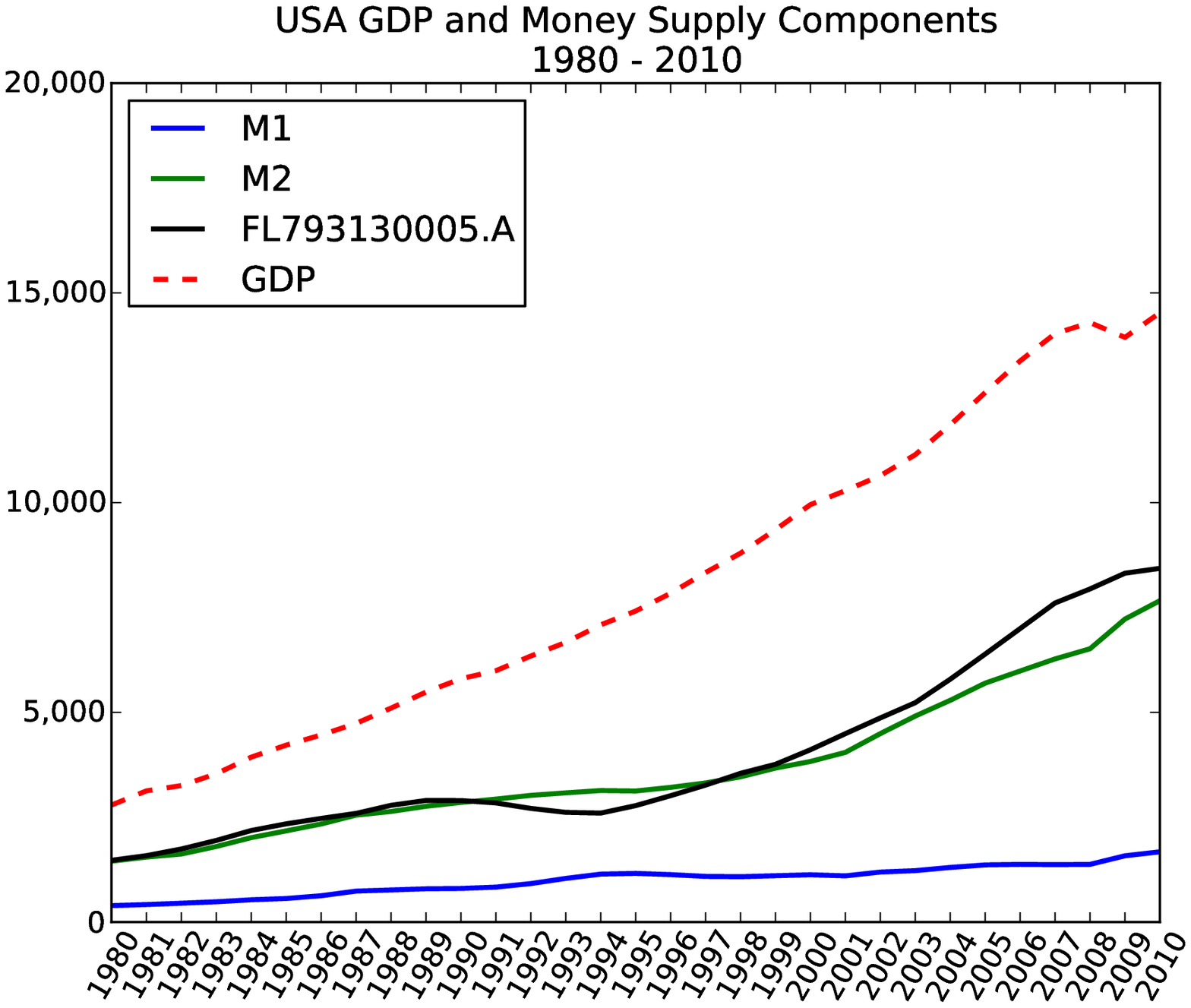}
\caption{GDP and Money Supply Components}
\label{fig:us_ms}
\end{center}
\end{minipage}
\hfill
\begin{minipage}[t]{7.5cm}
\begin{center}
\includegraphics[width=7.5cm, clip]{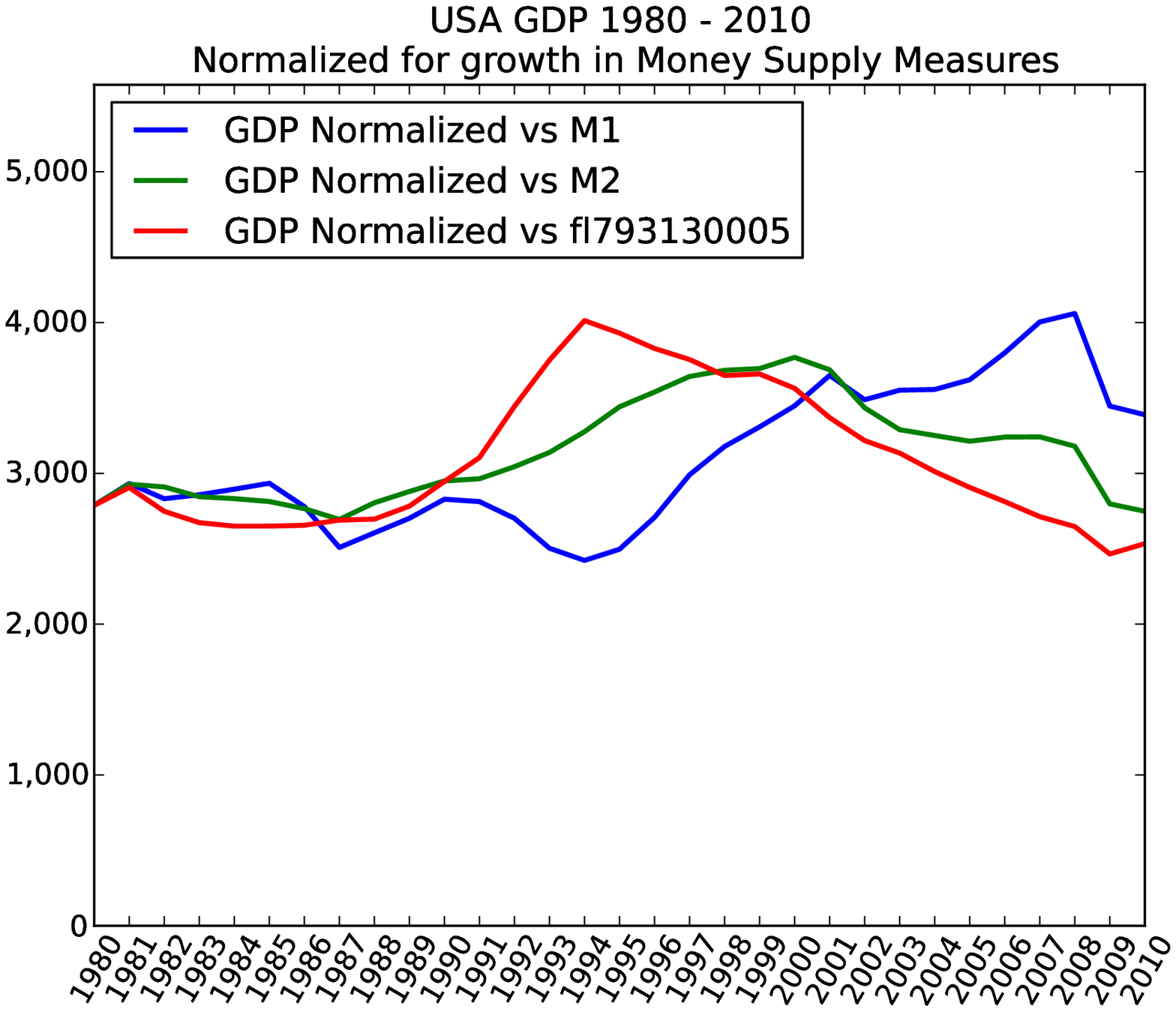}
\caption{Normalized US GDP} 
\label{fig:us_gdp}
\end{center}
\end{minipage}
\end{figure}
Statistics on the United States dollar are provided by the Board of 
Governors of the Federal Reserve 
System\footnote{\url{http://www.federalreserve.gov/econresdata/releases/statisticsdata.htm}}
primarily in the money stock measures H.6 table. 
Money supply growth ranges between 2-2.5 times per decade over
the period shown.  United States money supply figures are available for two 
aggregate measures, M1 and M2 covering US commercial banks and thrift savings
organisations. The M3 measure was discontinued in February 2006 and was significantly
contaminated with financial instruments representing debt. 
Seasonally and non-seasonally
adjusted values are available, the data shown here is from the non-seasonally
adjusted table. 
The M1 measure comprises physical currency, nonbank traveler's
cheques, demand and other chequeable deposits, and M2 consists of savings deposits,  
small-denomination time deposits, and retail money funds in addition to M1. Retail money funds 
have been excluded from the measures in the data shown as they are primarily composed of debt.
Large time deposits, which were a component of the M3 measure, are not included in the
M2 measure though. The Flow of Funds Report (Table Z.1), separately provides Table L.205
which also contains data on the financial system, and in particular the FL793130005.A table
which provides annual information on the total time and savings deposits in the financial
system, which we have included
for comparison. This data however, presumably does not include physical notes and coins, as can
be seen by its behaviour in the early 1990's, when it temporarily goes below M2. We suspect that
although it is probably a better measure of the total money supply than M2, it is still probably
not a complete one.
\par
Retail money funds are a form of money market fund offered 
primarily to individuals. Although money market funds are frequently described
as equivalent to money, table L.121 in the Federal Reserve Flow
of Funds data series provides a breakdown that shows that approximately 75\% of their
holdings are held in credit market instruments, for example Agency
and GSE backed securities. Their remaining funds appear to be held in different
forms of checkable deposit. There is no indication in the H.6 report that
these deposits are being excluded from the main total, leading to the suspicion that
inclusion of retail money funds in the report may also indicate double counting in addition to
the problematic classification of debt instruments as money.  
US money supply measures do not include the reserves held at the central bank, which are captured in
table H.3. These are currently significant due to the injection of funds by the Federal Reserve
as part of the TARP and Quantitative Easing interventions, and the payment of interest on these reserves.
\par
Closer examination
of the M1 measure shows that the physical component is growing at a slightly
faster rate than its other components, which may help explain the behaviour of the M1 normalised
GDP statistic. However,
a further confounding factor is the classification of accounts within the US
Banking system, where net-transaction accounts which are included under M1, 
require a reserve to be held at the Federal Reserve Banks. In practice, banks can control this
classification to some extent, resulting in funds being shifted between classification
under the M1 or M2 measure presumably
depending on the interest rate paid by the Federal Reserve on the reserve accounts held by the
banks at the Reserve. \footnote{Although Economic textbooks (see Mankiw\cite{mankiw.1997})
typically show reserves as being a fraction of deposits, they are shown on bank's balance sheets
as assets, rather than liabilites, implying that they represent additional funds, 
rather than a simple percentage of deposits.}
\par
The normalized M2 and FL793130005 GDP data is generally consistent with a monetary flow
shift into GDP measured items during the nineties, followed by a shift into
the financial sector during the first decade of the 21st century, but there are too many
other distortions in the US data to be definitive.
Indeed it is probably unwise, for the reasons provided above to draw too many conclusions
from the normalised GDP statistics for any country. However the USA is particularly
problematic, given the
difficulty of currently establishing the exact numbers for its effective money supply. 
Besides the issues presented by the Federal Reserve's statistics,  the United States is also
somewhat anomolous in its de facto position as global currency, with substantial
holdings of physical currency outside of the USA. 
\newpage
\subsection{European Union (Eurozone): 2002-2010}
\begin{figure}[ht]
\begin{minipage}[t]{7.5cm}
\begin{center}
\includegraphics[width=7.5cm, clip]{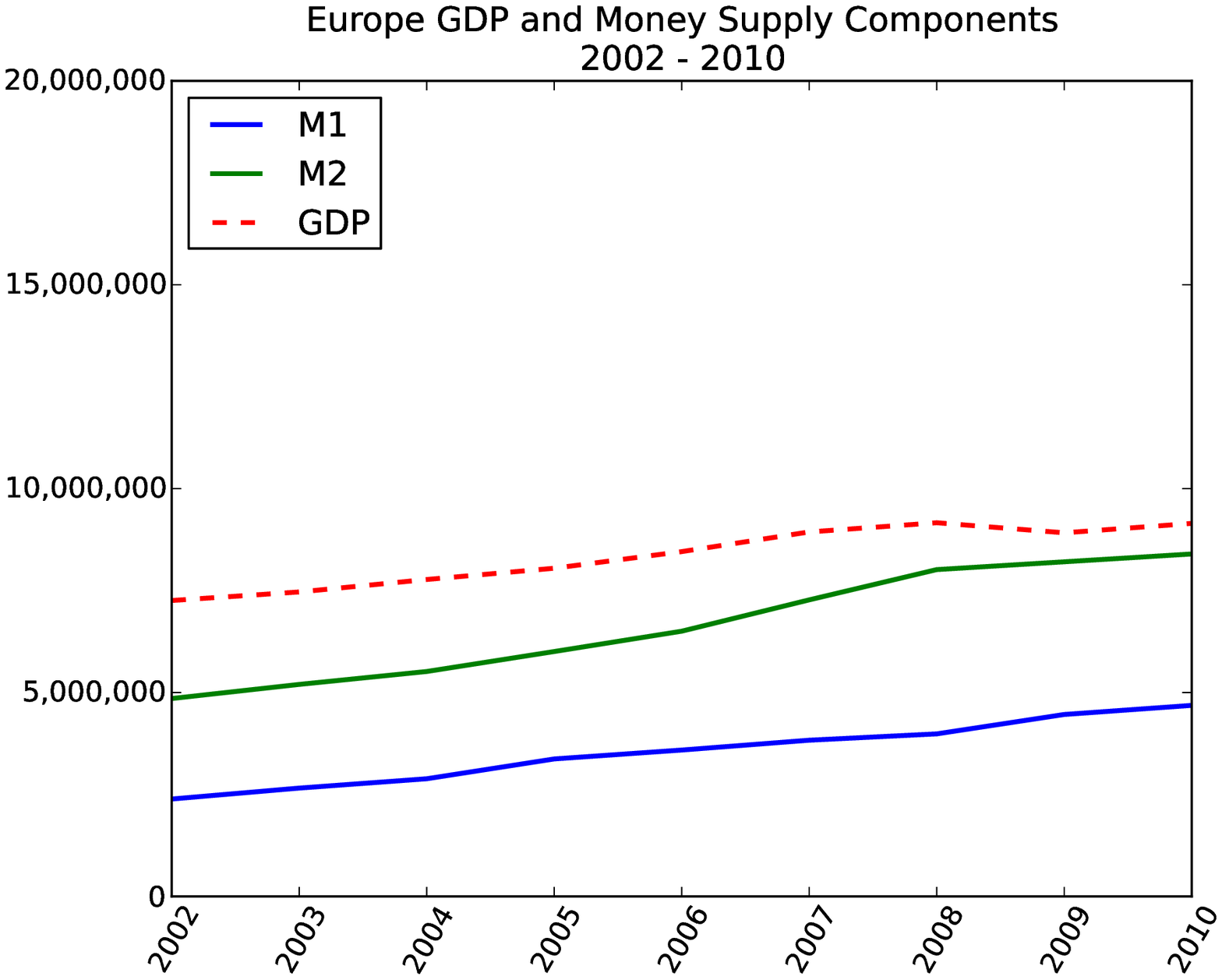}
\caption{GDP and Money Supply Components}
\label{fig:euro_ms}
\end{center}
\end{minipage}
\hfill
\begin{minipage}[t]{7.5cm}
\begin{center}
\includegraphics[width=7.5cm, clip]{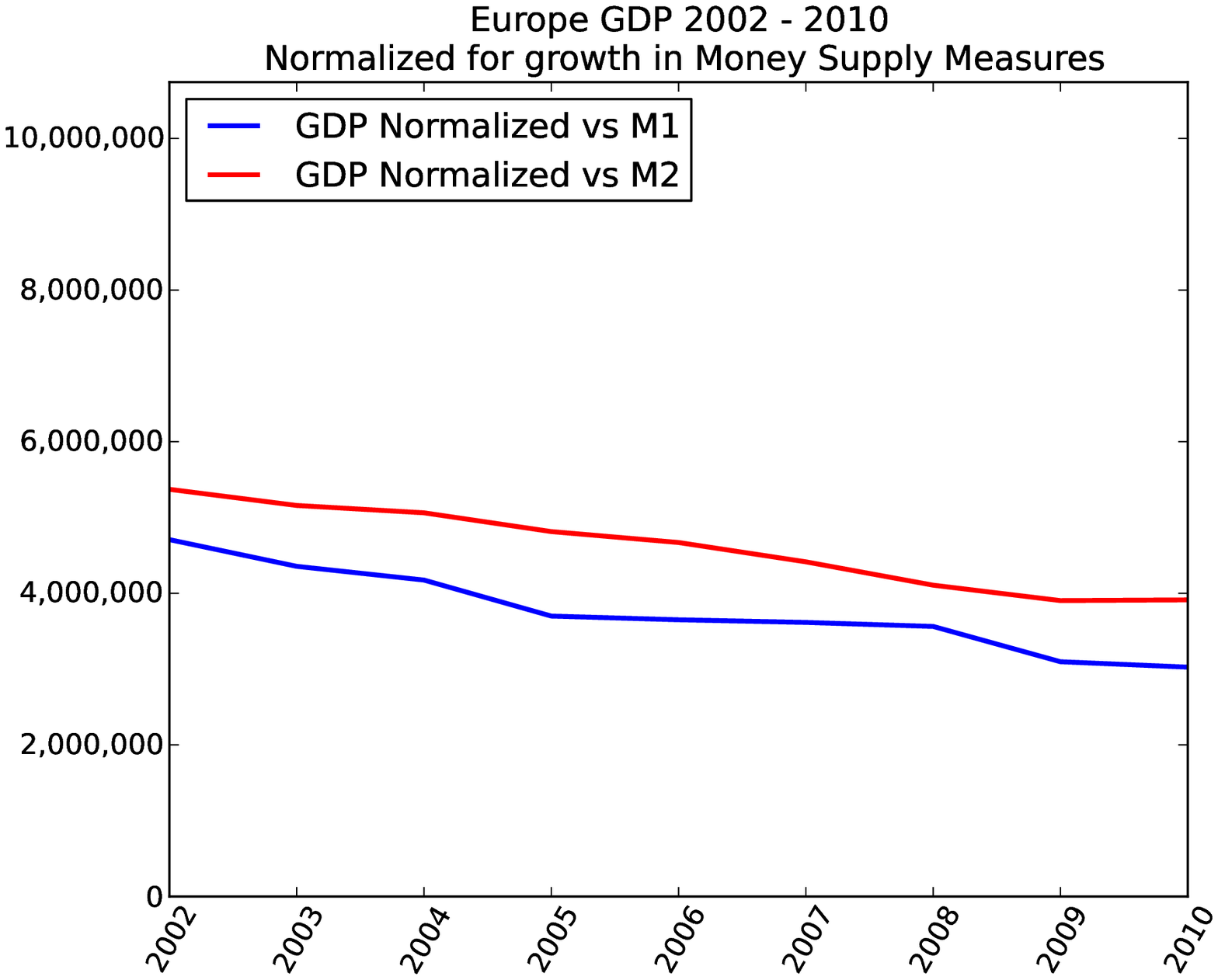}
\caption{Normalized Eurozone GDP 2002-2010}
\label{fig:euro_gdp}
\end{center}
\end{minipage}
\end{figure}
The Euro was established as a common currency for the European Union, with the exception then of
Denmark and the UK, on January 1st 1999. Creation was preceeded by several years of 
fiscal integration. Slovenia joined the Eurozone in 2007, Cyprus and Malta in 2008, and Slovakia
in 2009. Monetary aggregates for the entire zone are available from 1999, however
consolidated GDP data appears to be only available from 2002 from the European Commission's
statistical site, Eurostat\footnote{\url{http://epp.eurostat.ec.europa.eu/portal/page/portal/eurostat/home/}.
IMF data is available, but is denominated in dollars}.
\par
M1, M2, and M3 aggregates are available. M1 is defined as the sum of currency in circulation
and overnight deposits; M2 as M1 plus deposits with an agreed maturity of up to two years and deposits redeemable at notice of up to three months; and M3 is the sum of M2, repurchase agreements, money market fund 
shares/units and debt securities up to two years. We have not shown M3 here, since its additional
components are primarily composed of debt based financial instruments. 
\par
The Euro shows notable stability over the period of examination, with a total expansion rate of 1.96
between 2000-2010, slightly under that of the USA for the same period. Some qualifications should
be made both about the monetary aggregates and the GDP, since it highly probable that the consolidation
of these statistics is hiding significantly different rates of expansion and "growth", in the
member countries. Similar qualifications probably also apply to any currency used by large 
populations, in that significant regional differences can be expected over time, depending on the
behaviour of the local monetary system. The normalised behaviour of GDP measure trends
consistently down as predicted.

\subsection{United Kingdom: 1988 - 2010}
\begin{figure}[ht]
\begin{minipage}[t]{7.5cm}
\begin{center}
\includegraphics[width=7.5cm, clip]{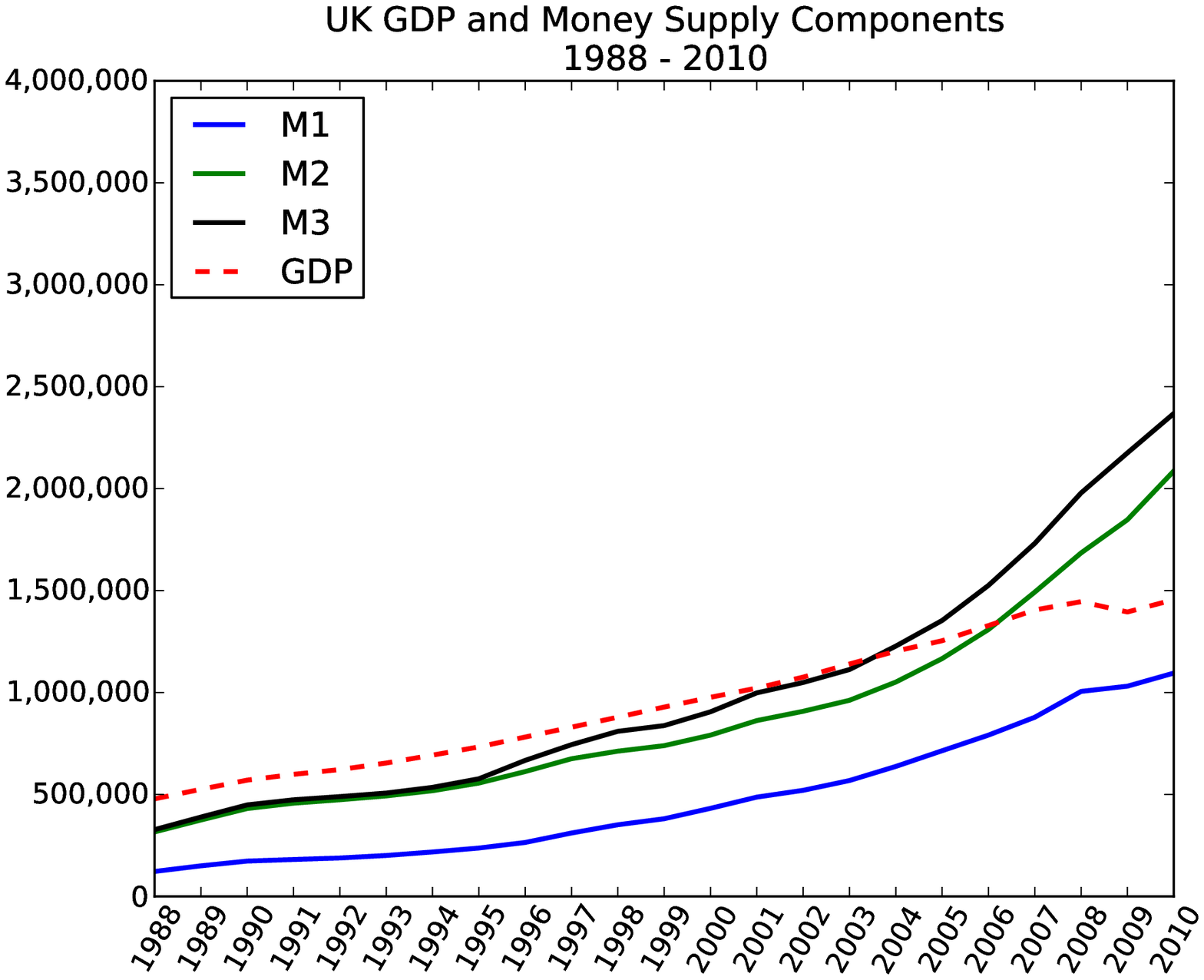}
\caption{GDP and Money Supply Components}
\label{fig:uk_ms}
\end{center}
\end{minipage}
\hfill
\begin{minipage}[t]{7.5cm}
\begin{center}
\includegraphics[width=7.5cm, clip]{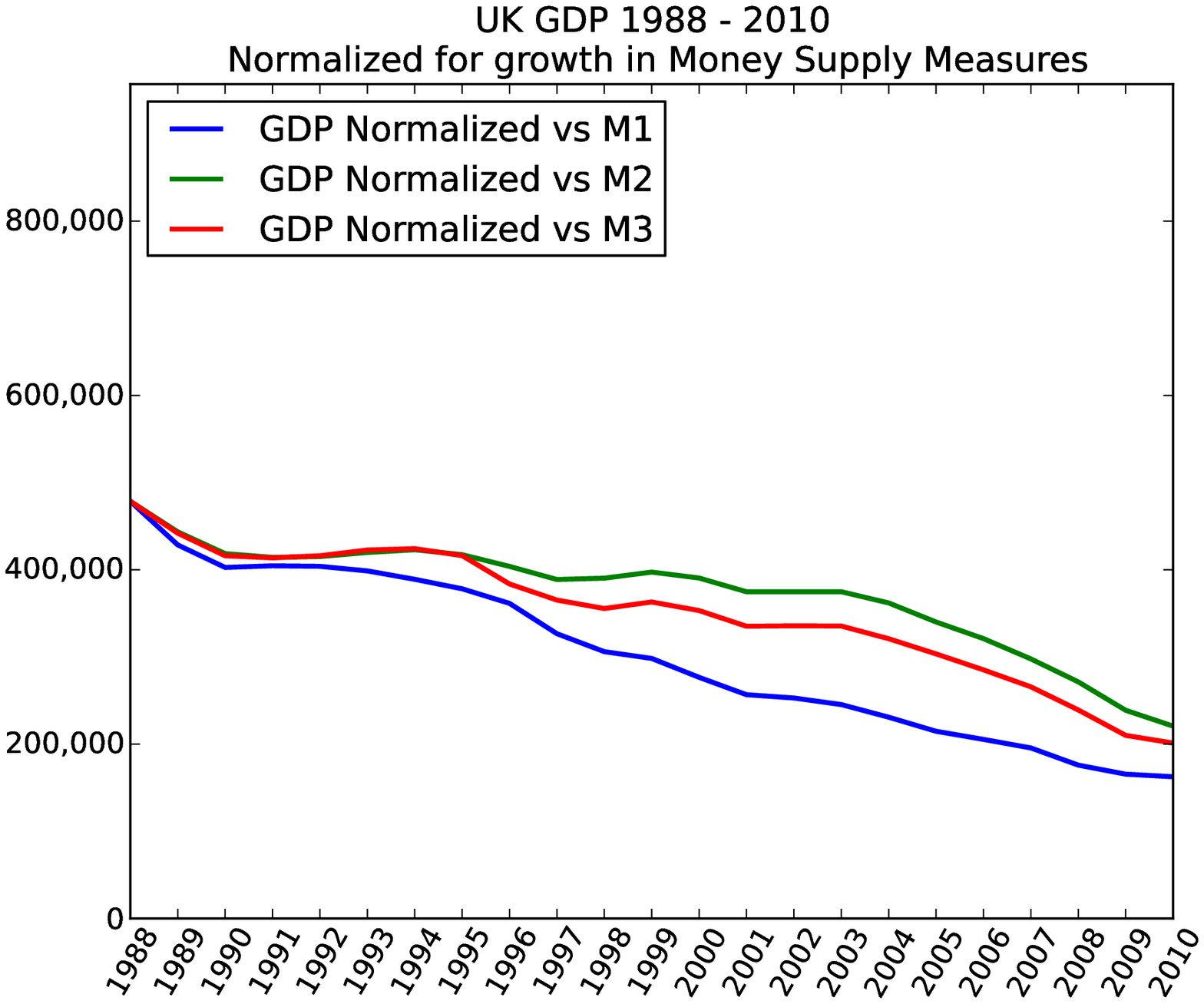}
\caption{Normalized UK GDP 1988-2010}
\label{fig:uk_gdp}
\end{center}
\end{minipage}
\end{figure}
Data for the United Kingdom is provided by the Bank of 
England\footnote{\url{http://www.bankofengland.co.uk/statistics/index.htm}} and is
available from 1988. Money supply growth ranges from 1.9 - 2.6 times per decade 
over the period of measurement.
The Bank of England publishes its money supply figures in the form
of its own M4 statistic, which appears to be an unfortunate mixture of deposits,
currency and forms of debt such as 'repos', commercial paper 
and bonds\footnote{Repurchase agreements (repos) are financial instruments representing
the sale of securities, with an agreement by the seller to repurchase the securities
at a later date, at typically a higher price. They are in effect a form of
collaterised debt. Commercial paper is an unsecured promissory note, with a fixed
maturity of 1 to 270 days.} However, M1, M2 and M3 statistics are also provided
as part of a monthly series of data submitted to the European Central Bank, and 
these are shown in Figure \ref{fig:uk_ms} although an exact description of the 
components of these series is not provided.\footnote{Source tables from the Bank
of England's interactive statistical database are M1: LPMVWYT, M2: LPMVWYW and 
M3: LPMVWYZ.} 
\par
Interestingly growth in the GDP measure for the UK is somewhat less than that for 
the underlying money supply components over the last 10 years. Sterling is
also somewhat anomolous in that it was, circa 1900 the previously dominant
global currency, and is still internationally significant as 
a consequence of the UK's position as a global financial centre with financial 
trading dominating the UK monetary system in comparison to GDP measured production. 
The steady decrease in the normalized GDP measurements shown in Figure \ref{fig:uk_gdp} 
is suggestive of 
a steady shift of money into the financial sector over the period shown.
\subsection{Switzerland: 1985 - 2010}
\begin{figure}[ht]
\begin{minipage}[t]{7.5cm}
\begin{center}
\includegraphics[width=7.5cm, clip]{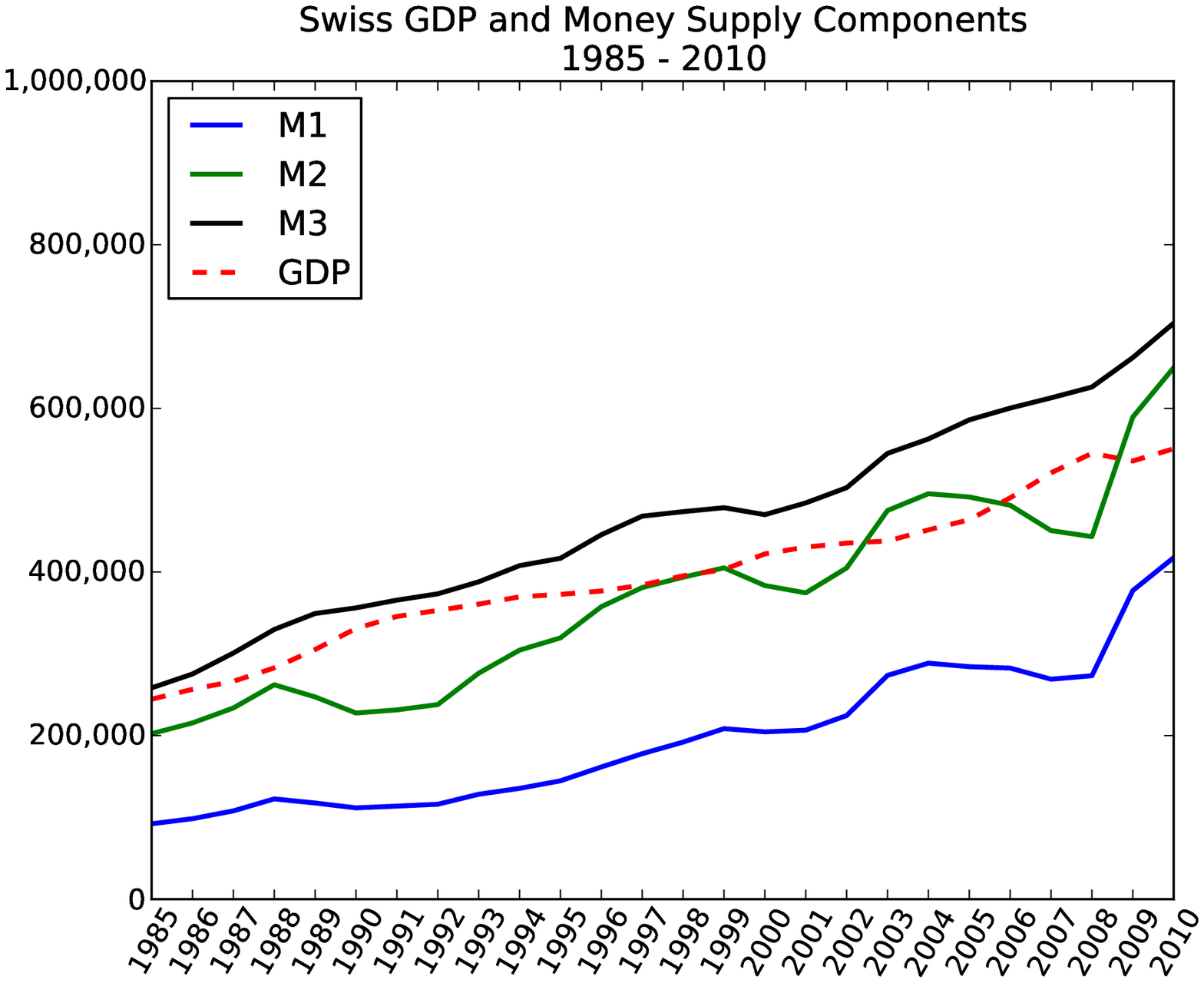}
\caption{GDP and Money Supply Components}
\label{fig:swiss_ms}
\end{center}
\end{minipage}
\hfill
\begin{minipage}[t]{7.5cm}
\begin{center}
\includegraphics[width=7.5cm, clip]{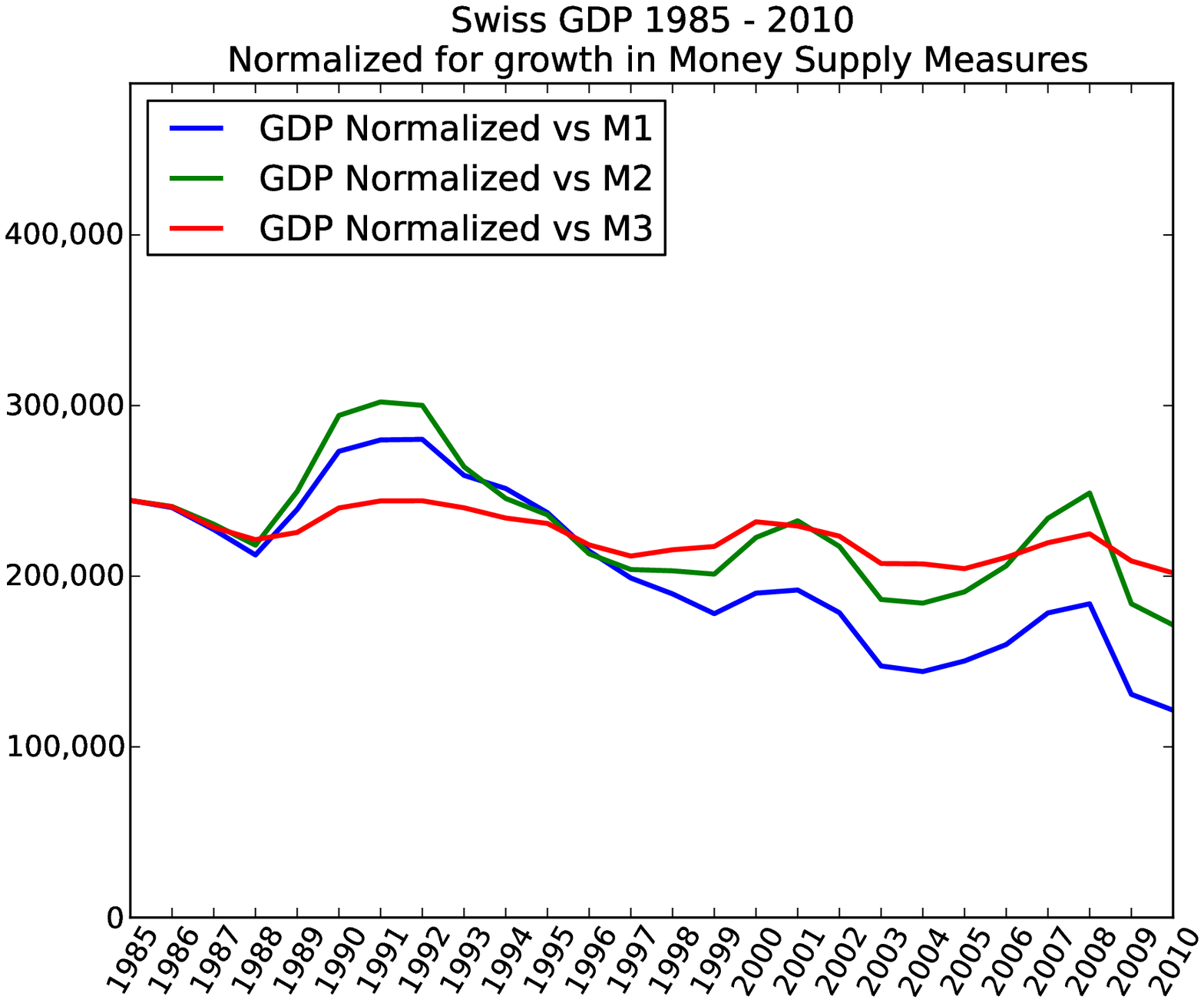}
\caption{Normalized Swiss GDP 1985-2010}
\label{fig:swiss_gdp}
\end{center}
\end{minipage}
\end{figure}
Data for Switzerland is available from the Swiss National 
Bank\footnote{\url{http://www.snb.ch/}} and includes the Duchy of Liechenstein. 
Figures are provided for the M1, M2 and M3
monetary aggregates. M1 is defined as Currency in Circulation, sight deposits
and deposits in transaction accounts, M2 as Savings deposits plus M1, and M3 as
time deposits plus M2. The behaviour of the M1 and M2 measures in comparison
with the M3 measure of total money supply shown in Figure \ref{fig:swiss_ms} suggests that
money is being shifted within the banking system between accounts classified
under the different measures, and this is reflected in the behaviour of the normalised
measures. A small monetary contraction appears to have also occurred in 2000.
The Swiss Franc exhibits one of the
lowers growth rates of the currencies being examined, wth the M3 measure growing
by 1.7 times in the 2000-2010 period. 
\par
\subsection{Japan 1985-2010}
\begin{figure}[ht]
\begin{minipage}[t]{7.5cm}
\begin{center}
\includegraphics[width=7.5cm, clip]{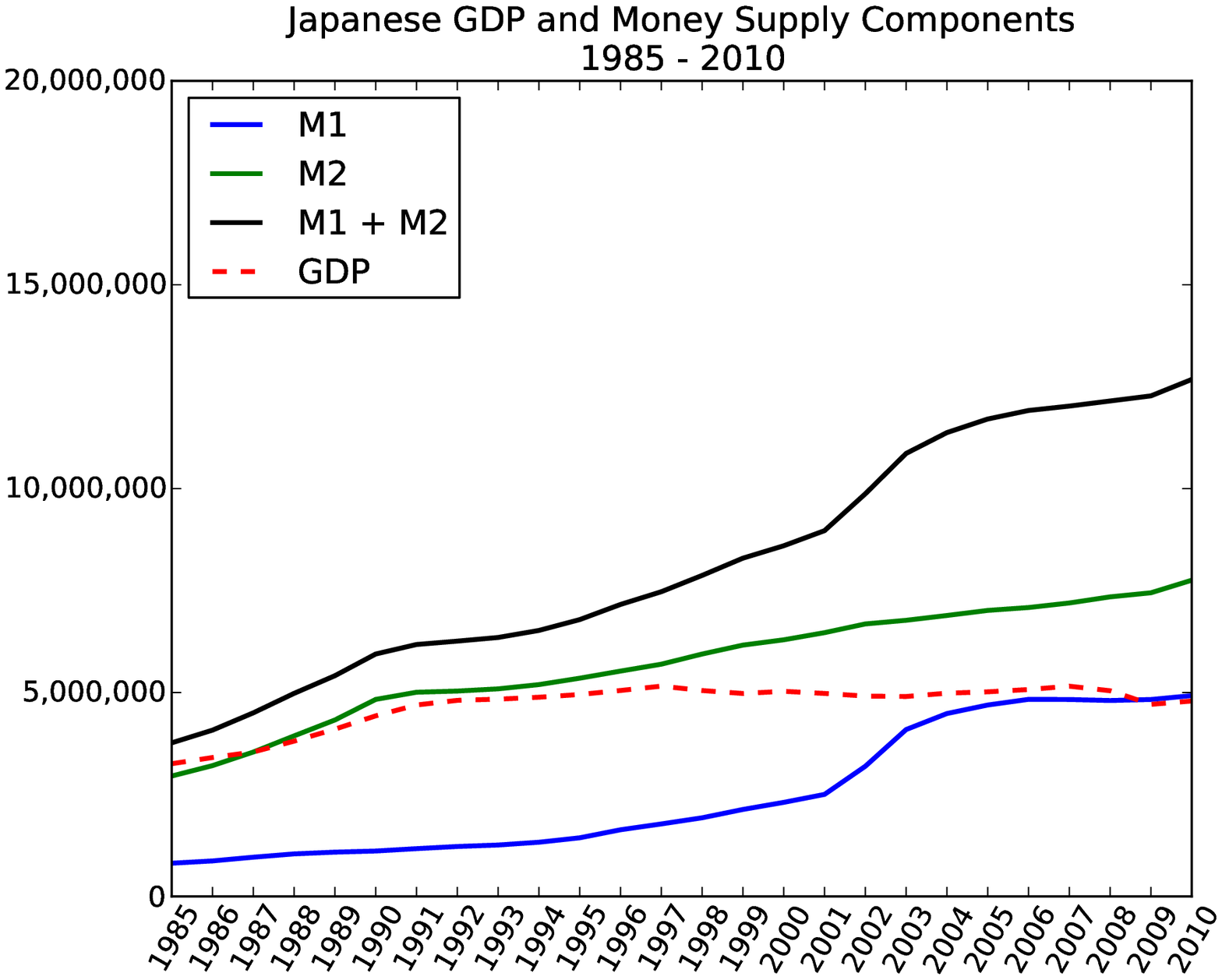}
\caption{GDP and Money Supply Components}
\label{fig:japanese_ms}
\end{center}
\end{minipage}
\hfill
\begin{minipage}[t]{7.5cm}
\begin{center}
\includegraphics[width=7.5cm, clip]{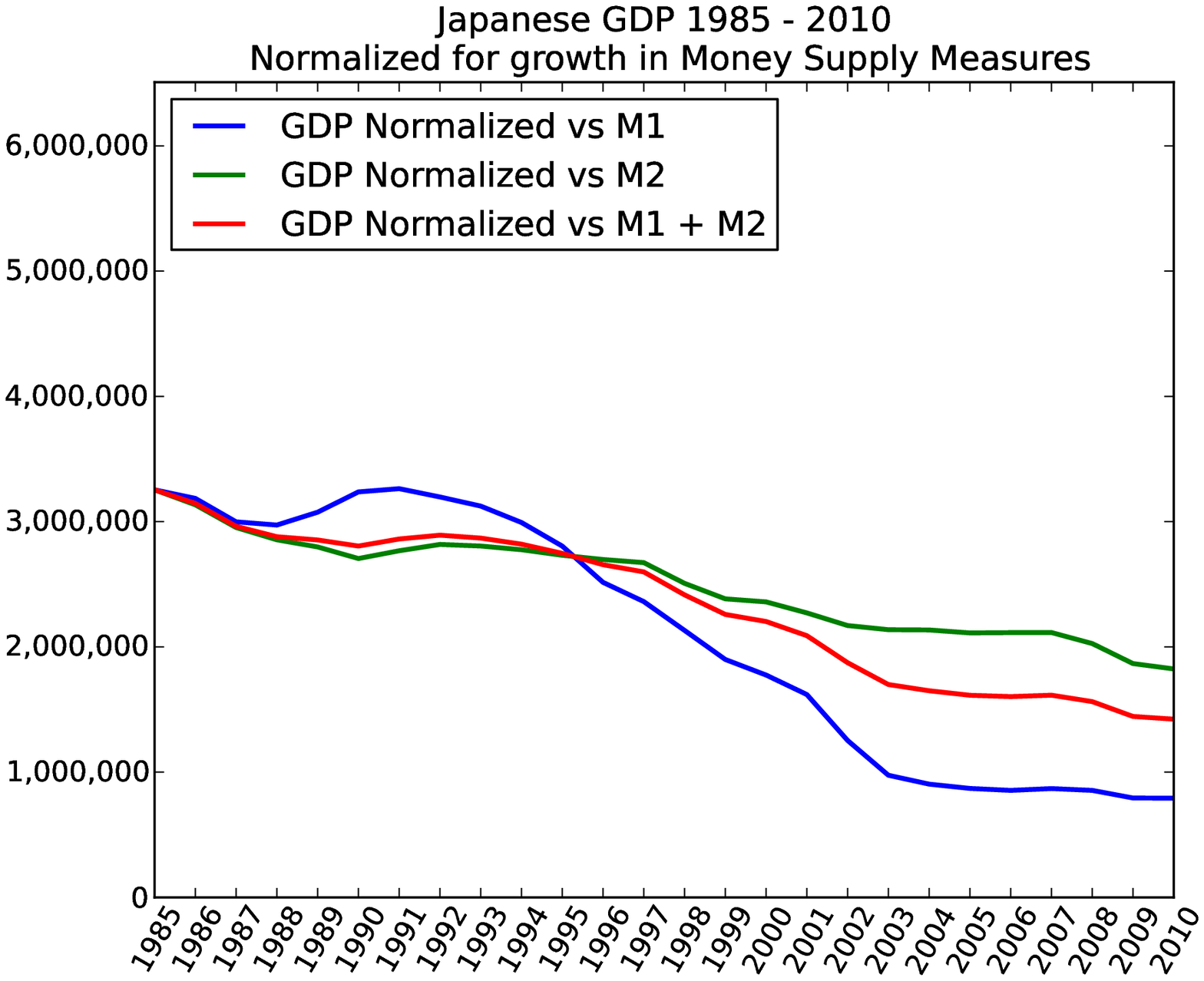}
\caption{Normalized Japanese GDP 1985-2010}
\label{fig:japanese_gdp}
\end{center}
\end{minipage}
\end{figure}
Data for Japan is available from the Bank of Japan's statistical 
database\footnote{\url{http://www.stat-search.boj.or.jp/index_en.html}}
The Bank of Japan lists a total of 4 money stock indices, M1, M2, M3 and L.
M1 is defined as currency in circulation and deposit money, where deposit
money is defined as demand deposits (current deposits, ordinary deposits,
savings deposits, deposits at notice, special deposits and deposits
for tax payments) minus checks and bills held by the depsitory institutions.
M2 has the same definition but extends the range of institutions to 
include the Bank of Japan, foreign banks, domestically licensed banks, and the
Shinkin Banks
which are small regional
financial institutions similar in many respects to US credit unions. 
Unusually, the M2 definition does not include M1, and excludes             
the Japan Post Bank which is a major domestic financial institution,
which is included in M1.  We have consequently shown M1, M2 and M1 + M2 separately.
\par
M3 is defined as M1 plus quasi-money and certificates of deposit. Quasi
money appears to include debt, government and bank bonds, and bank 
debentures.\footnote{There is some discrepancy between the English
and Japanese descriptions with the Japanese description providing a more detailed
enumeration.} L is then further defined as a range of debt instruments,
commercial paper, government securities and foreign bonds plus M3.
Owing to the presence of significant proportions of debt instruments in the M3
and L measures, we have elected to exclude them from the data presented here.
\par
The data itself is broken up into three separate series in the Bank of 
Japan's statistical database as shown in Table \ref{tab:japan_m}, 
which have been concatenated for this analysis.
\begin{table}[ht]
\centering
\begin{tabular}{c|c|c}
Period    &  M1 & M2 \\
\hline
1985-1998 &  MA'MAMS1AN01 & MA'MAMS1ANM2C \\
1998-2003 &  MA'MAMS3AN01 & MA'MAMS3ANM2C \\
2003-2010 &  MA'MAMS5ANM1 & MA'MAMS5ANM2  \\
\hline
\end{tabular}
\caption{Bank of Japan Data Series References}
\label{tab:japan_m}
\end{table}
\par
Over the period of measurement Japan has generally experienced an extremely
low rate of monetary expansion, which is reflected in the behaviour of both its normalised
and un-normalised GDP growth. The growth rate for the Japanese Yen over the period 2000-2010
for example is 1.5 compared with 2x for the US dollar, however markedly higher rates
of expansion seem to have occurred during the Japanese credit bubble which ended in 1990.
\subsection{India}
\begin{figure}[ht]
\begin{minipage}[t]{7.5cm}
\begin{center}
\includegraphics[width=7.5cm, clip]{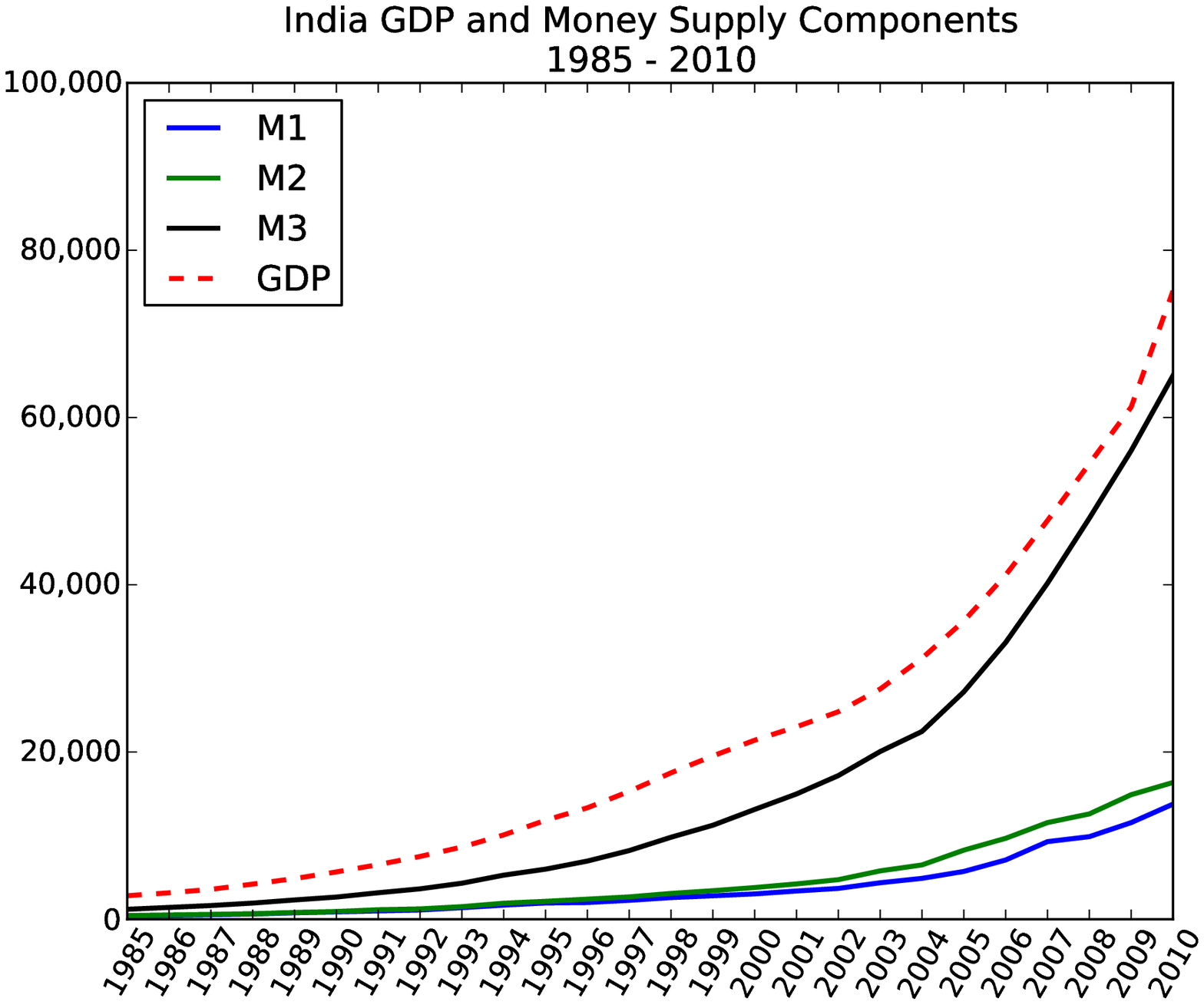}
\caption{GDP and Money Supply Components}
\label{fig:india_ms}
\end{center}
\end{minipage}
\hfill
\begin{minipage}[t]{7.5cm}
\begin{center}
\includegraphics[width=7.5cm, clip]{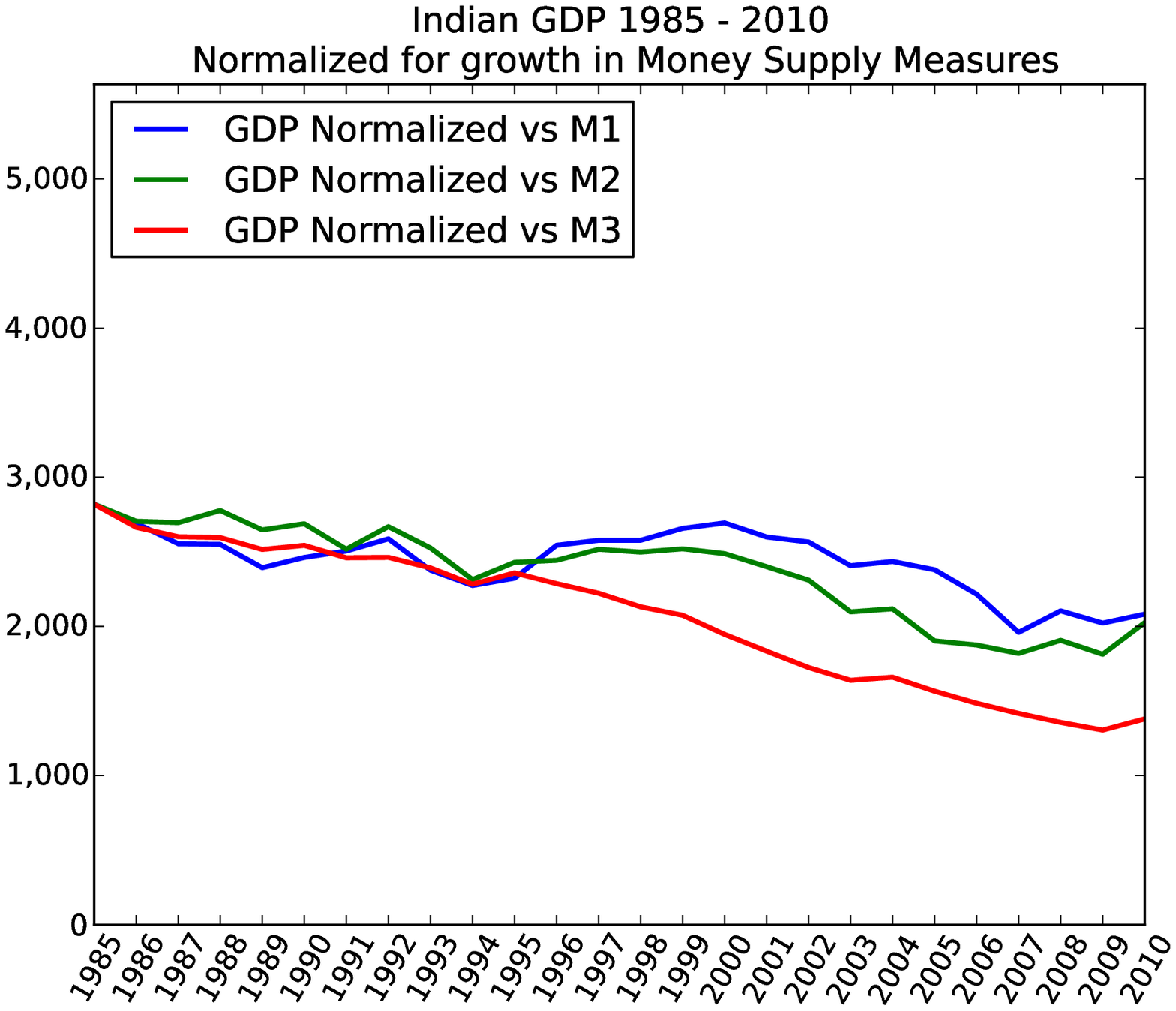}
\caption{Normalized Indian GDP 1985-2010}
\label{fig:india_gdp}
\end{center}
\end{minipage}
\end{figure}
Information on the Indian Rupee's money supply is available from the Reserve Bank of 
India\footnote{\url{http://www.rbi.org.in/scripts/statistics.aspx}}. India shows a high
rate of growth over the period shown, with an average expansion rate for M3 of 5 times per decade.
Indian money supply figures are available for three aggregate measures, M1, M2, and M3. 
The Indian definition of M1 is Currency with the Public, Current Deposits with the Banking System, 
Demand Liabilities Portion of Savings Deposits with the Banking System, 'Other' Deposits with the 
RBI\footnote{http://rbidocs.rbi.org.in/rdocs/PublicationReport/Pdfs/32272.pdf}
M2 is defined as M1 plus Currency with the Public, Current Deposits with the Banking System plus 
Savings Deposits with the Banking System, Certificates of Deposit issued by Banks, Term 
Deposits of residents with a contractual maturity up to and including one year with the 
Banking System (excluding CDs) and 'Other' Deposits with the RBI
M3 is defined as M2 plus Term Deposits of residents with a contractual maturity of over one year 
with the Banking System and Call/Term borrowings from 'Non-depository' Financial Corporations by the 
Banking System. 
Figure \ref{fig:india_ms} shows the raw  values for the three money supply components and GDP since 1985.
\par
GDP data normalized separately against all 3 money supply components is shown in 
Figure \ref{fig:india_gdp}. The behaviour of the 2 lower normalised components in comparison with 
M3 is again suggestive of money being shifted between differently classified accounts, rather
than any providing any economic meaning. Although the overall behaviour is also suggestive
of money being shifted into the financial sector, in the Indian context, it is probably also
indicative of the increasing provision of banking services in the Indian economy to larger
numbers of its citizens.
\subsection{China: 2000 - 2010}
\begin{figure}[ht]
\begin{minipage}[t]{7.5cm}
\begin{center}
\includegraphics[width=7.5cm, clip]{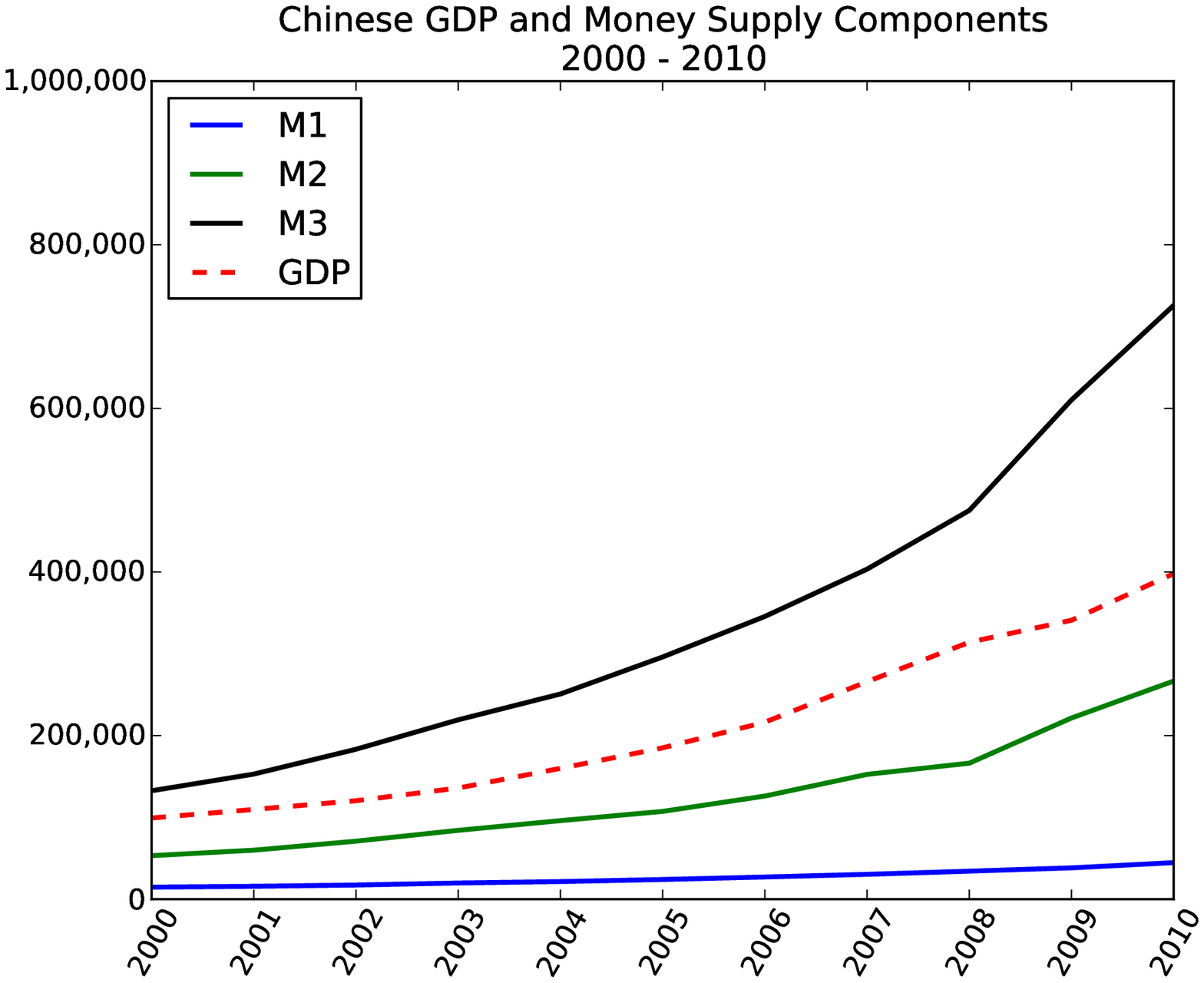}
\caption{GDP and Money Supply Components}
\label{fig:chinese_ms}
\end{center}
\end{minipage}
\hfill
\begin{minipage}[t]{7.5cm}
\begin{center}
\includegraphics[width=7.5cm, clip]{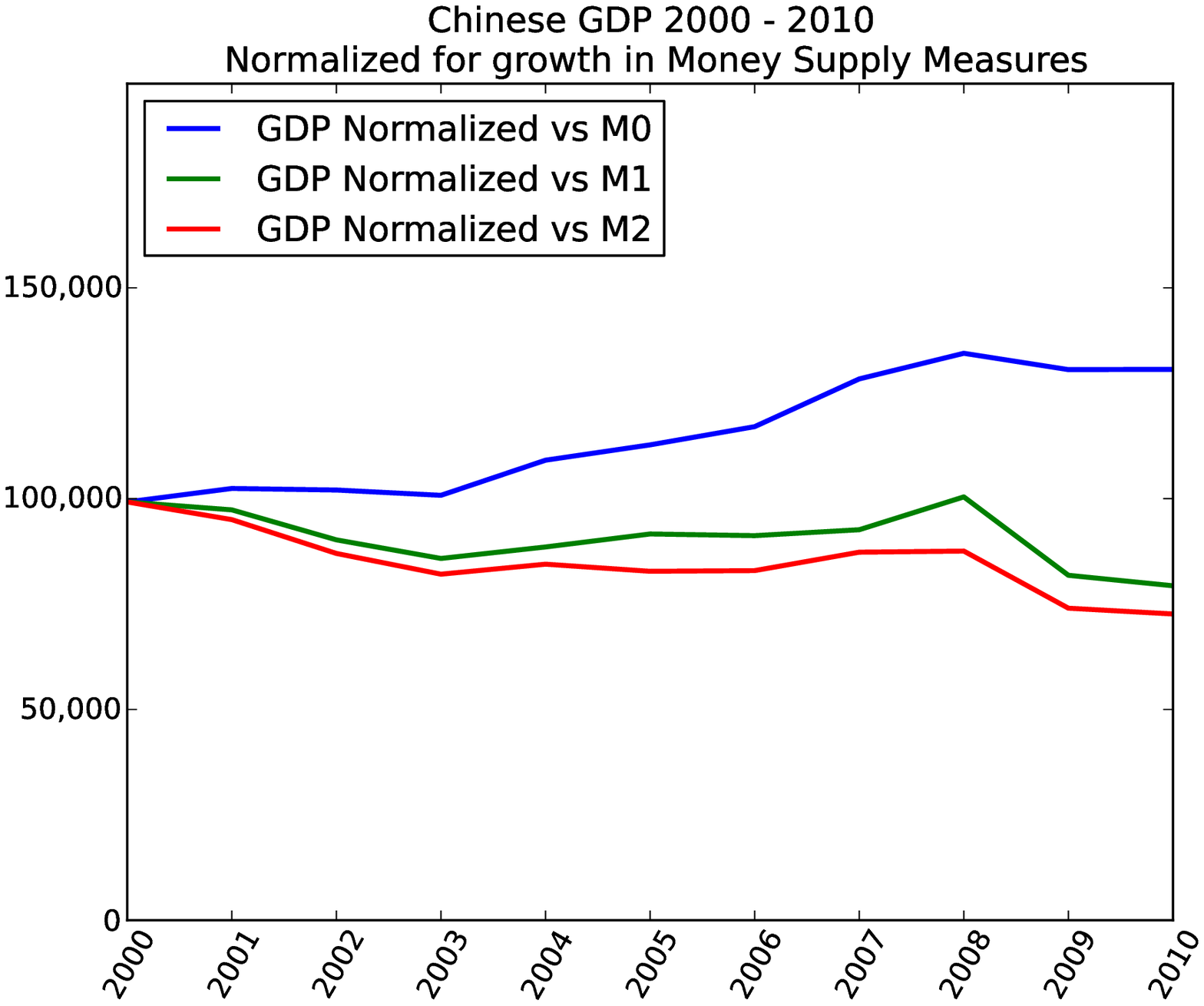}
\caption{Normalized Chinese GDP }
\label{fig:chinese_gdp}
\end{center}
\end{minipage}
\end{figure}
Chinese money supply measures are available for three aggregate measures, M0, M1, and M2 
from the People's Bank of China\footnote{\url{http://www.pbc.gov.cn/publish/english/963/index.html}} 
for the period 2000-2010 under the Monetary Survey statistical series. On the more 
detailed series provided in 2003, M0 is listed as Currency in Circulation, M1 as the sum of
M0 and Demand Deposits, and M2 as the sum of M0, M1 and Time, Savings and Other deposits. 
\par
Figure \ref{fig:chinese_ms} shows the data for the three money supply components with
the GDP figures from the IMF, Gross Domestic Product current prices, national currency
series. On average over the entire period, the M0 monetary measure increased by 3 times,
whilst the M1 and M2 measures by slightly over 5 times. Similarly to the situation in India, 
the difference in the relative
increase of physical and non-physical money suggests a movement from physical to 
non-physical(bank deposit) currency over the period, as well as an increase in the size of
the financial sector of the economy.
\subsection{Iceland 1990 - 2010}
\begin{figure}[ht]
\begin{minipage}[t]{7.5cm}
\begin{center}
\includegraphics[width=7.5cm, clip]{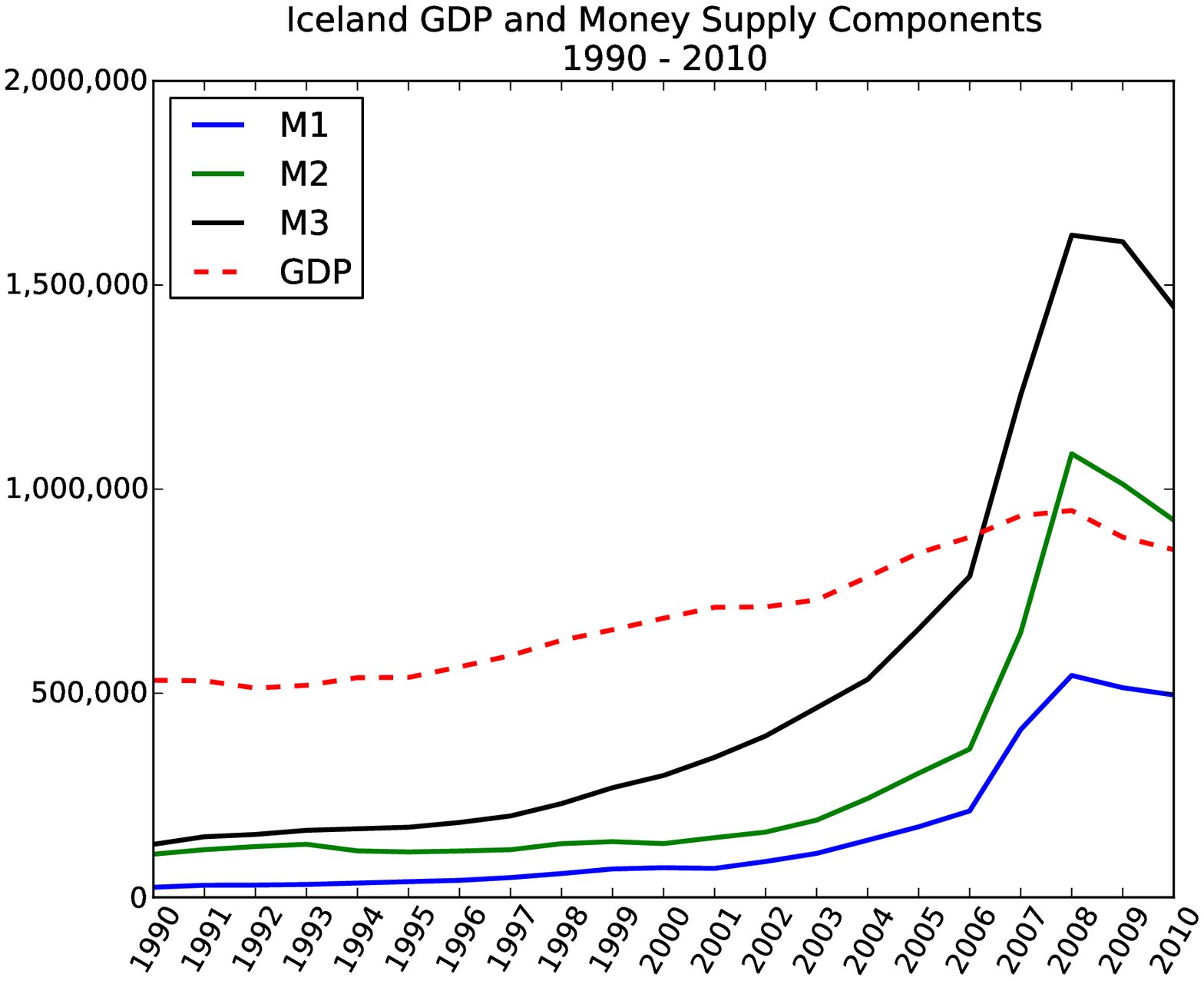}
\caption{GDP and Money Supply Components}
\label{fig:iceland_ms}
\end{center}
\end{minipage}
\hfill
\begin{minipage}[t]{7.5cm}
\begin{center}
\includegraphics[width=7.5cm, clip]{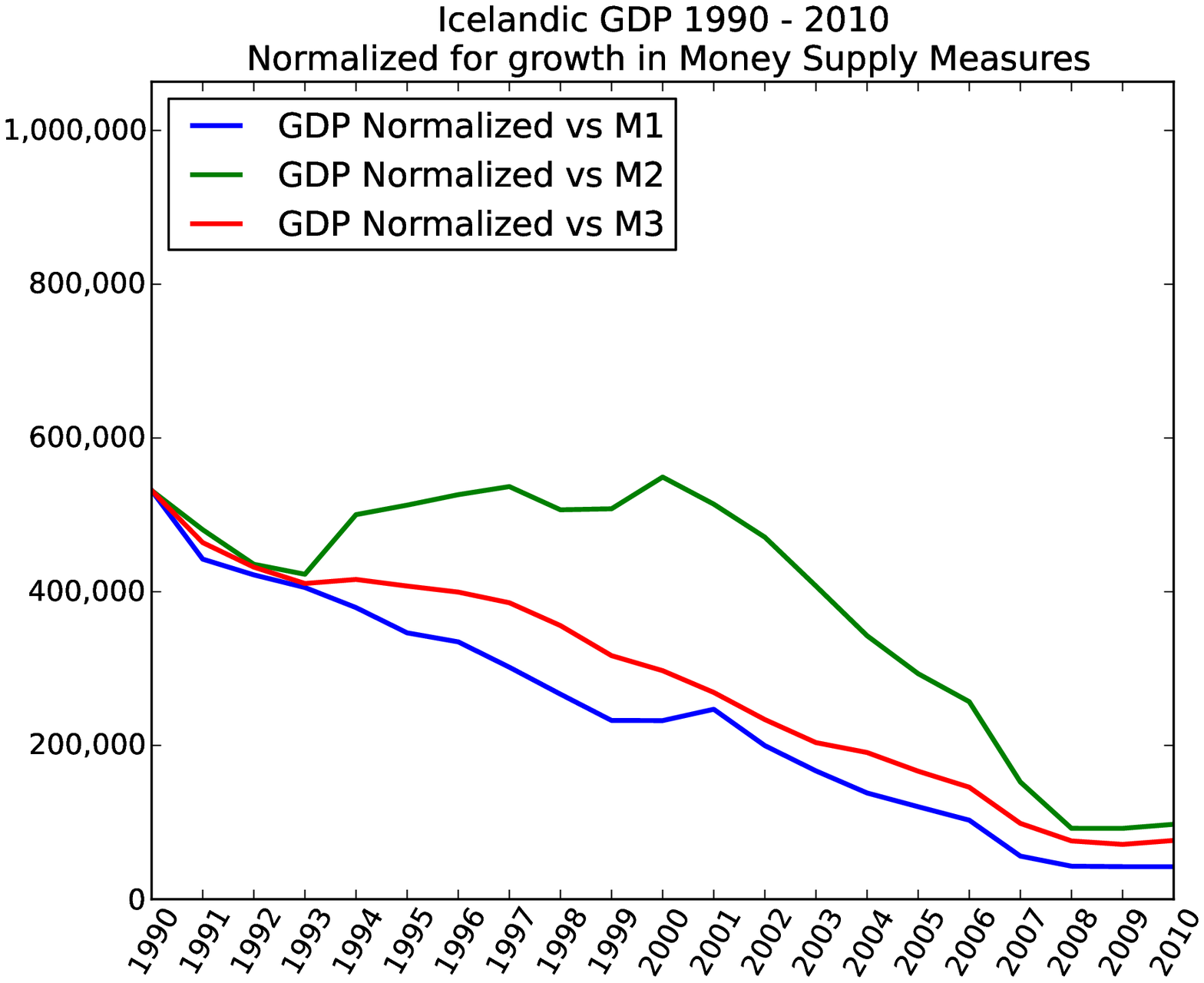}
\caption{Normalized Icelandic GDP }
\label{fig:iceland_gdp}
\end{center}
\end{minipage}
\end{figure}
Data on the Icelandic Money Supply is available from the Icelandic Central 
Bank\footnote{\url{http://www.sedlabanki.is/?pageid=194}}. Three statistical
series are provided, M1, M2 and M3, where M1 is defined as physical notes
and coins and demand deposits, M2 as sight deposits plus M1, and M3 as time deposits
plus M2. Iceland is the smallest economy examined (national population is 320,000),
but is shown here since it is often used as a reference for economic success, based
on the performance of its GDP.
\par
Time series are published under the monetary statistics heading of statistics
section on the central bank's web site. However, recent releases have removed
data before 2010 from this series. The bank's annual report, which can be found
under publications on the same web site\footnote{\url{http://www.sedlabanki.is/?PageID=235}}
also includes the detailed money supply statistics in its appendix for the year of 
publication and the preceding 
6 years. Reports from 1997 to 2010 can be accessed online, and access to printed
copies preceding that can be obtained by application to the central bank's library.
Data presented in this analysis is obtained from time series data downloaded
before data was removed in September 2010, and by cross-reference with the Annual
Report. A number of small inconsistencies were encountered, both between 
different annual reports and versions of the downloadable data suggesting that
the Icelandic central bank appears to be engaged in a process of continuous revisionism, but
the size of the discrepancies themselves is too small to be significant for this analysis.
\par
Iceland experienced an extreme expansion of its credit and money supplies during the 2000's,
with its M3 money supply growing by 10 times over the period, in contrast to 2.3 times
during the 1990's
and this can be seen clearly in the money supply expansion, and subsequent partial
contraction during the banking system collapse. Prior to the 1990's however, Iceland
had experienced hyper-inflation for much of the 1970's and 1980's.
The expansion in the 2000's was also accompanied by   
a considerable increase in the financial sector, 
and this can be seen clearly in the normalised statistics. 
\subsection{New Zealand: 1988 - 2010}
\begin{figure}[ht]
\begin{minipage}[t]{7.5cm}
\begin{center}
\includegraphics[width=7.5cm, clip]{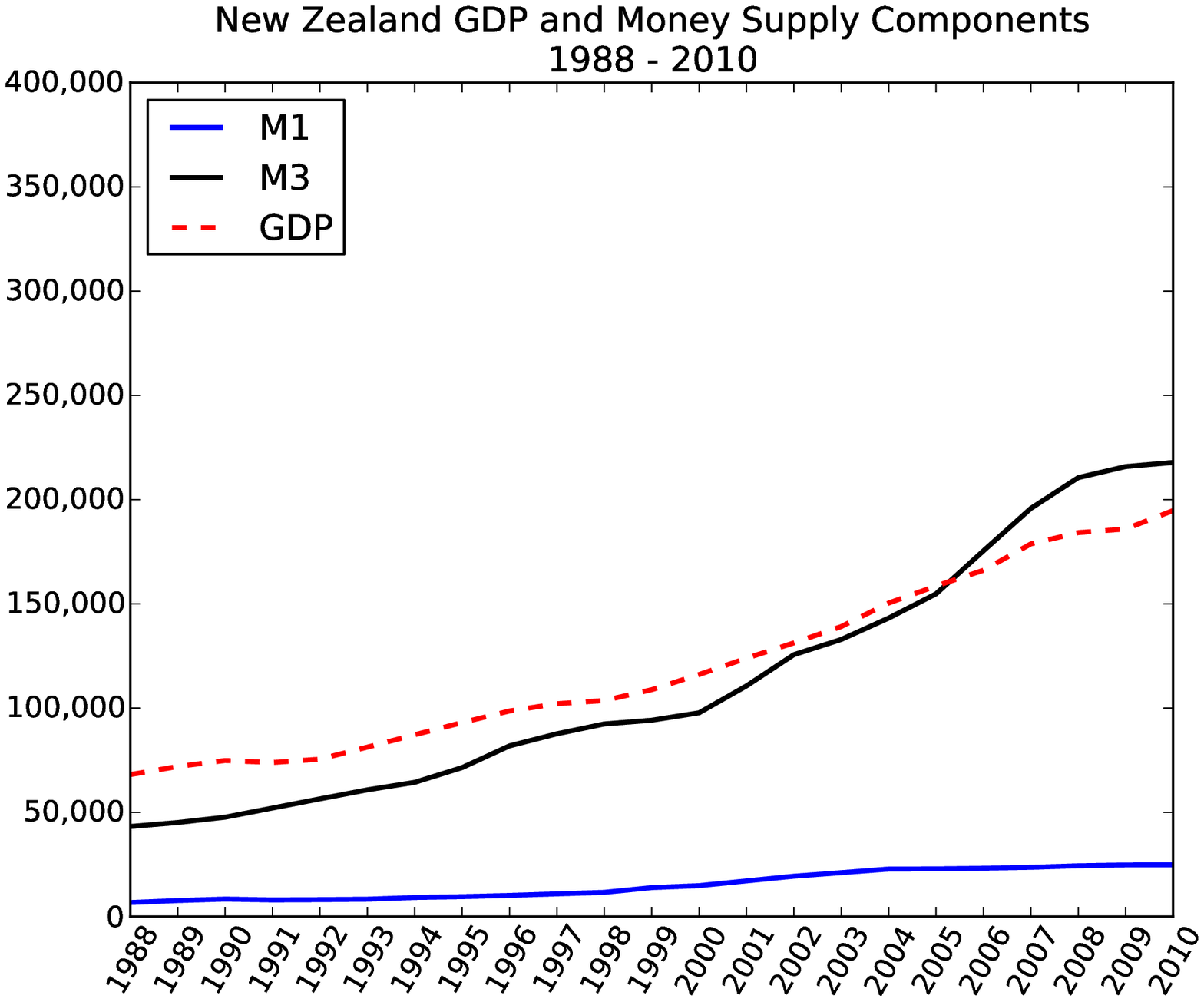}
\caption{GDP and Money Supply Components}
\label{fig:new_zealand_ms}
\end{center}
\end{minipage}
\hfill
\begin{minipage}[t]{7.5cm}
\begin{center}
\includegraphics[width=7.5cm, clip]{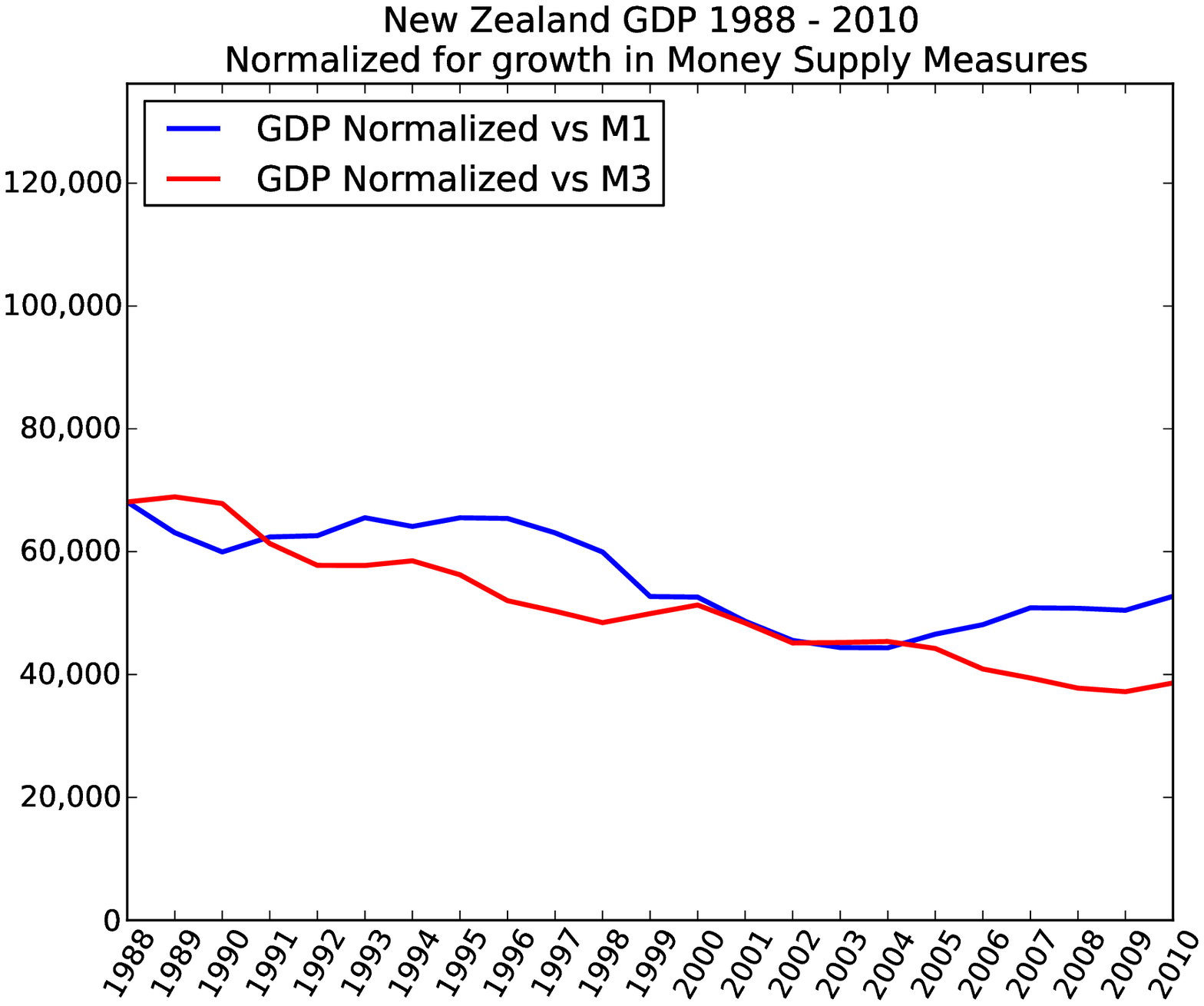}
\caption{Normalized New Zealand GDP }
\label{fig:new_zealand_gdp}
\end{center}
\end{minipage}
\end{figure}
Information on New Zealand's money supply is available from the Reserve Bank of New 
Zealand\footnote{\url{http://www.rbnz.govt.nz/statistics/monfin/c1/description.html}}
Several aggregate measures are published, including M1, M2, and several variations of
M3. M1 is defined as physical notes and coins, plus chequeable deposits, minus
inter-institutional chequeable deposits and central government deposits. M2
is defined as M1 plus all non-M1 call funding, and M3 is M1 plus M2 plus NZ dollar
funding minus inter-M3 institutional claims. Other versions of M3 exlude
repurchase agreements and funding from non-residents.
\par
New Zealand is included here as an example of an advanced predominently agricultural 
economy, with a relatively small financial sector, although it has experienced a
significant housing bubble in the last few years.
Data for the M1 and M3 raw measures is presented here, since both the M2 and M3 data appears
to include some forms of debt in the form of overnight funding arrangements. 
On average the New Zealand money supply expanded by approximately 2 times
per decade over the period examined.
Behaviour
of both of the normalised GDP measurements is though consistent with the hypotheses
advanced here. 
\subsection{Russia: 1997 - 2010}
\begin{figure}[ht]
\begin{minipage}[t]{7.5cm}
\begin{center}
\includegraphics[width=7.5cm, clip]{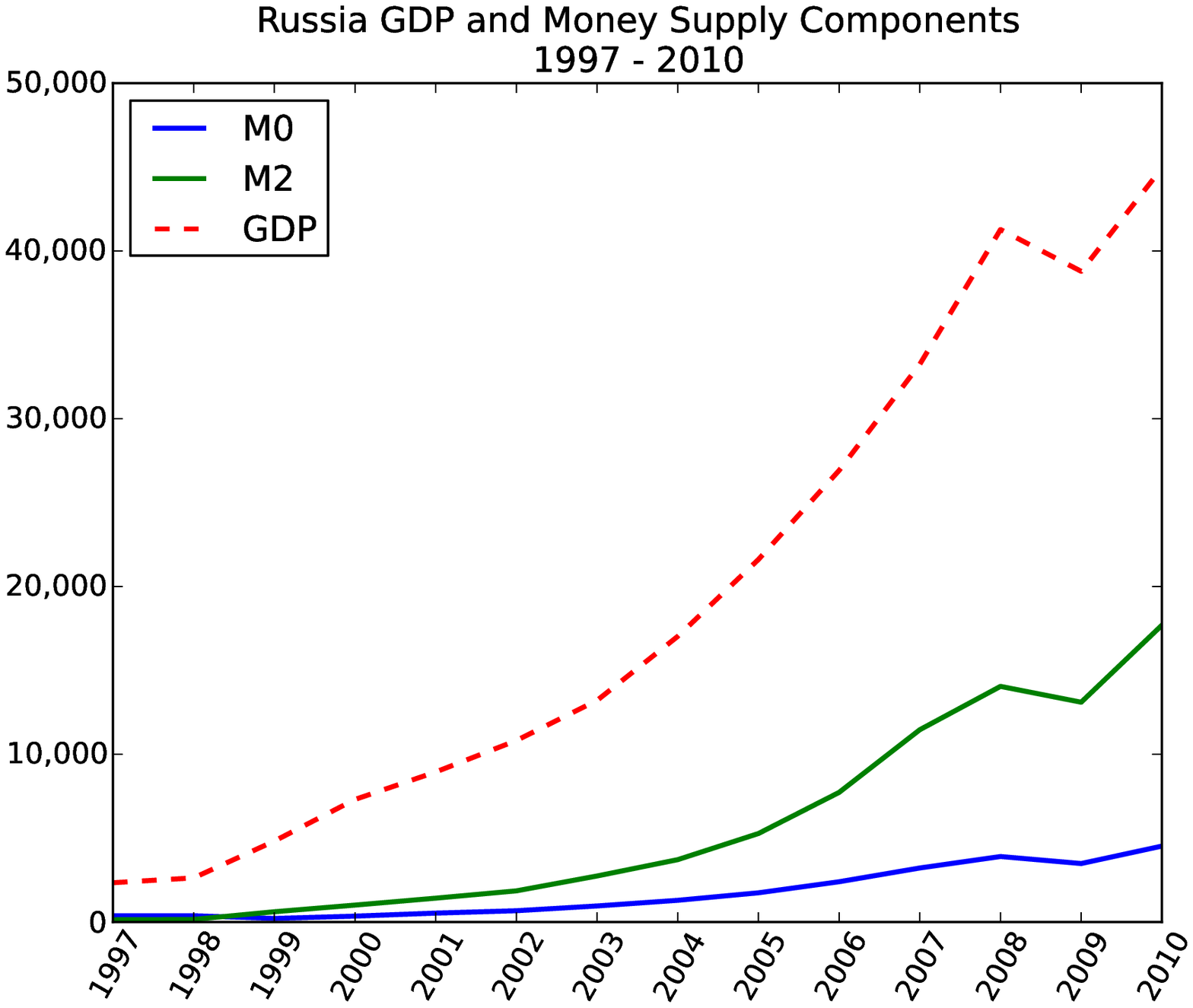}
\caption{GDP and Money Supply Components}
\label{fig:russia_ms}
\end{center}
\end{minipage}
\hfill
\begin{minipage}[t]{7.5cm}
\begin{center}
\includegraphics[width=7.5cm, clip]{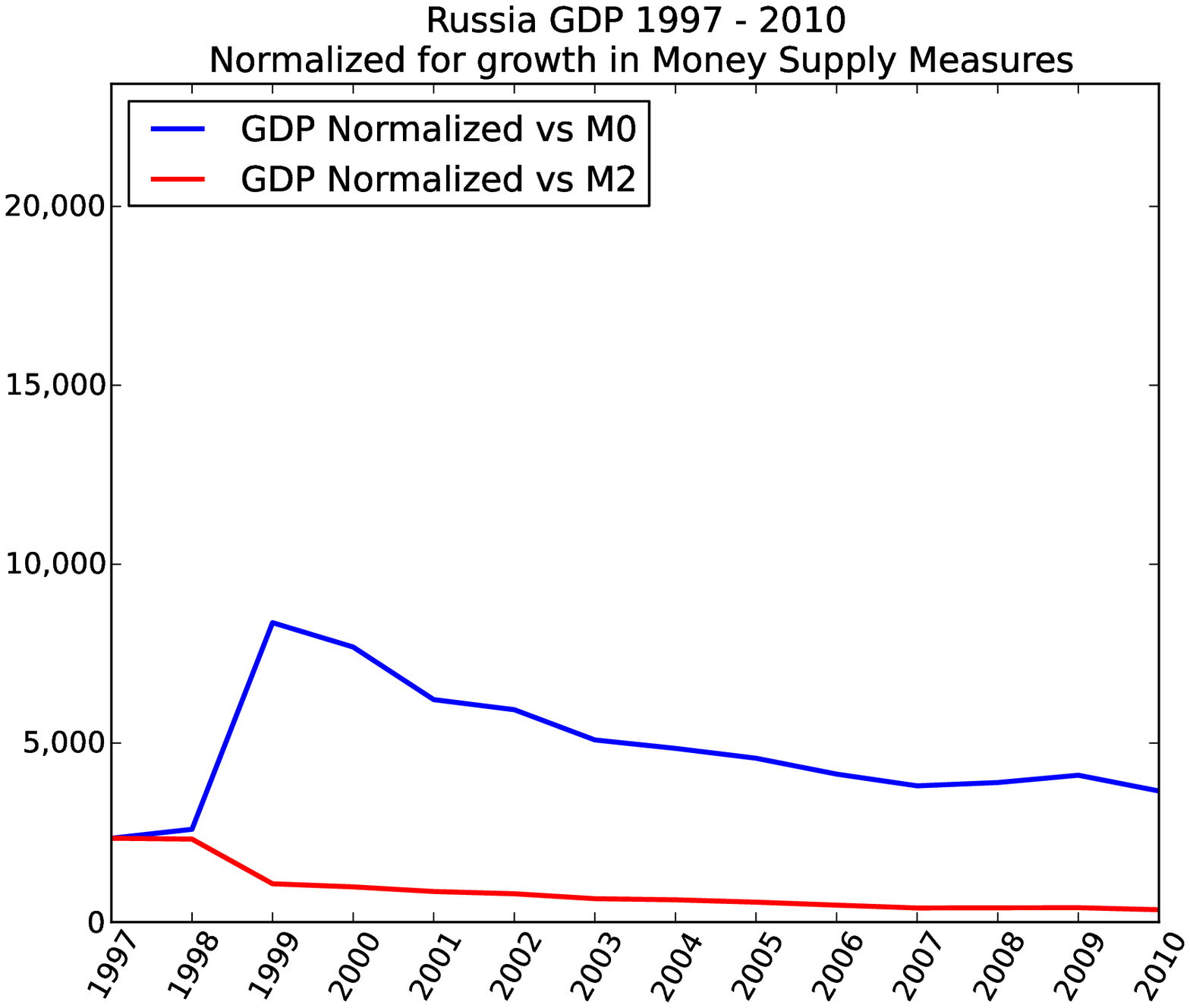}
\caption{Normalized Russian GDP }
\label{fig:russia_gdp}
\end{center}
\end{minipage}
\end{figure}

Data on the Russian money supply can be obtained from the Central Bank of 
Russia\footnote{\url{http://www.cbr.ru/eng/statistics/credit_statistics/MS.asp?Year=2011}} which
provides M0 and M2 measures, where M0 is defined as physical
notes and coins, and M2 as M0 plus all non-cash funds of resident non-financial
and financial institutions (except for credit ones) and private individuals in rubles. This
definition appears to exactly meet the definition of money proposed here (physical notes
and coins, and all deposits in the banking system), and the monetary data is clearly presented 
on the web site, including monthly percentage changes in the different measures. 
\par
Russia is of interest both as a predominently energy and natural resources exporter, but
also as an example of a country with a recently reset monetary system.
Russia has had to repeatedly redenominate its currency over the last hundred years, with the 
latest reset occurring on 1st January 
1998, when 1 new ruble replaced 1000 of the previous denomination, after a period
of hyperinflation that followed the collapse of the Soviet Union in 1991. Since the redenomination,
significant quantitative inflation has continued, with an expansion of the M2 money supply between
2000-2010 of 18 times.

%% file: biblio.tex
\bibliography{finance}
\raggedright
\bibliographystyle{unsrt}